\pdfoutput=1
\documentclass[review,3p, number,sort&compress,11pt]{elsarticle}
\usepackage{graphicx,subcaption}
\usepackage{amsthm,amsmath,amssymb,amsfonts}
\usepackage{upgreek}
\theoremstyle{plain}

\theoremstyle{definition}

\usepackage{algorithmicx,algorithm}
\usepackage{algpseudocode}
\usepackage{tabularx}
\usepackage{multirow}
\usepackage[bottom]{footmisc}
\usepackage[skip=2pt,font=small]{caption}
\usepackage{multicol}
\usepackage{todonotes}
\usepackage{lipsum}
\usepackage{url}
\usepackage{enumitem}
\setlist{topsep=0pt,parsep=0pt, leftmargin=*}

\usepackage[T1]{fontenc}
\usepackage[utf8]{inputenc}

\newcommand{\ignore}[1]{}

\usepackage[nolist]{acronym}

\usepackage{verbatim} 

\usepackage[right]{eurosym} % €
\usepackage{amsmath, amssymb} % mathematische Symbole / mathematical symbols
\usepackage{mathtools}
\usepackage{gensymb} % degree symbols

\usepackage{graphicx}
\usepackage{subcaption} % allows subplots
\usepackage{pdfpages} % add PDF
\usepackage{adjustbox} % allows rotation

\usepackage{booktabs} % nicer tables
\usepackage{longtable} % 
\usepackage{tabularx} % additional settings
\usepackage{multirow} % connect cells
\usepackage{threeparttable} % other type of tables
\usepackage{float}
\usepackage{colortbl}
\usepackage{xcolor}
\usepackage{makecell}

\begin{acronym}[Bash]
	
	\acro{HAR}{human activity recognition}
	\acro{KD}{Knowledge Distillation}
	\acro{CE}{Cross Entropy}
	\acro{CF}{Catastrophic Forgetting}
	\acro{FC}{Fully Connected}
	\acro{DPP}{Determinantal point process}
	\acro{iCaRL}{Incremental Classifier and Representation Learning}
	\acro{GEM}{Gradient Episodic Memory}
	\acro{A-GEM}{Averaged-GEM}
	\acro{FIM}{Fisher-Information Matrix}
	\acro{LPL}{Locality Preserving Loss}
	\acro{FWSR}{Frank-Wolfe Sparse Representation}
    \acro{MAS}{Memory Aware Synapses}
    \acro{ILOS}{Incremental Learning In Online Scenario}
    \acro{WA}{Weight Alignment}
\end{acronym}

\usepackage{lineno}

\renewcommand\vec{\mathbf}

\journal{Information Sciences}

\begin{document}

\begin{frontmatter}
% https://www.overleaf.com/project/5ea04d00749d0500015ad630
% 30 pages
\title{Continual Learning in Sensor-based Human Activity Recognition: an Empirical Benchmark Analysis}



\author[label1]{Saurav Jha}
\author[label1]{Martin Schiemer}
\author[label2]{Franco Zambonelli}
\author[label1]{Juan Ye}
\address[label1]{School of Computer Science, University of St Andrews, UK}
\address[label2]{Dipartimento di Scienze e Metodi dell'Ingegneria, Universita' di Modena e Reggio Emilia, Italy}

\begin{abstract}
Sensor-based human activity recognition (HAR), i.e., the ability to discover human daily activity patterns from wearable or embedded sensors, is a key enabler for many real-world applications in smart homes, personal healthcare, and urban planning. However, with an increasing number of applications being deployed, an important question arises: how can a HAR system autonomously learn new activities over a long period of time without being re-engineered from scratch? This problem is known as continual learning and has been particularly popular in the domain of computer vision, where several techniques to attack it have been developed. This paper aims to assess to what extent such continual learning techniques can be applied to the HAR domain. To this end, we propose a general framework to evaluate the performance of such techniques on various types of commonly used HAR datasets. We then present a comprehensive empirical analysis of their computational cost and  effectiveness of tackling HAR-specific challenges (i.e., sensor noise and labels’ scarcity). The presented results uncover useful insights on their applicability and suggest future research directions for HAR systems.
% investigate the impact of in-memory sample size and training data size on the accuracy of these techniques.

\end{abstract}


\end{frontmatter}

\section{Introduction}
% define HAR 
Sensor-based human activity recognition (\ac{HAR}) is to infer human daily activities from data collected on various types of sensors~\cite{LIU201641}. It can be regarded as a \textit{classification} problem; that is, given raw sensor data, extracting features and classifying them into a class label; \textit{i.e.}, an activity. For example, by monitoring users' interaction with everyday objects via ambient binary sensors we can infer whether the user is watching TV or preparing a meal. By tracking users' movement via accelerometers or gyroscopes on wearables we can recognise their physical activities such as jogging or climbing stairs. These daily activities can have a significant impact on a wide range of real-world applications~\cite{ALVAREZALVAREZ2013162}, ranging from smart home~\cite{8645807} and adaptive environments~\cite{ViroliZ10,Ahmed5255} to personal healthcare~\cite{8706961,MROZEK2020132,7737670} and disease diagnosis~\cite{AFONSO2019282}. 

We have witnessed an increasing number of HAR applications being deployed and running over longer spans~\cite{8903481,LIU201713}. This drives an important research question: \textit{how does a HAR system continuously discover and learn new types of activities}? We cannot assume that once a HAR system is trained with an initial set of activities, then the users of the system will only perform the same set of activities all the time. People tend to change their behaviour patterns for internal or external factors~\cite{ROS201386}. Such change requires a HAR system to adapt their learning and include new types of activities in order to provide desired services. For example, the COVID outbreak has impacted the routine of people across the world, including practising different exercises, cooking new cuisines, and switching work patterns. For a personal healthcare monitoring application, failure to recognise new exercise routines may lead to misdiagnosis and inappropriate medication prescription. Therefore, continual learning is the key enabler for a long-term, sustainable HAR system. 

\begin{figure}[!htp]
\centering
\includegraphics[width=0.8\textwidth]{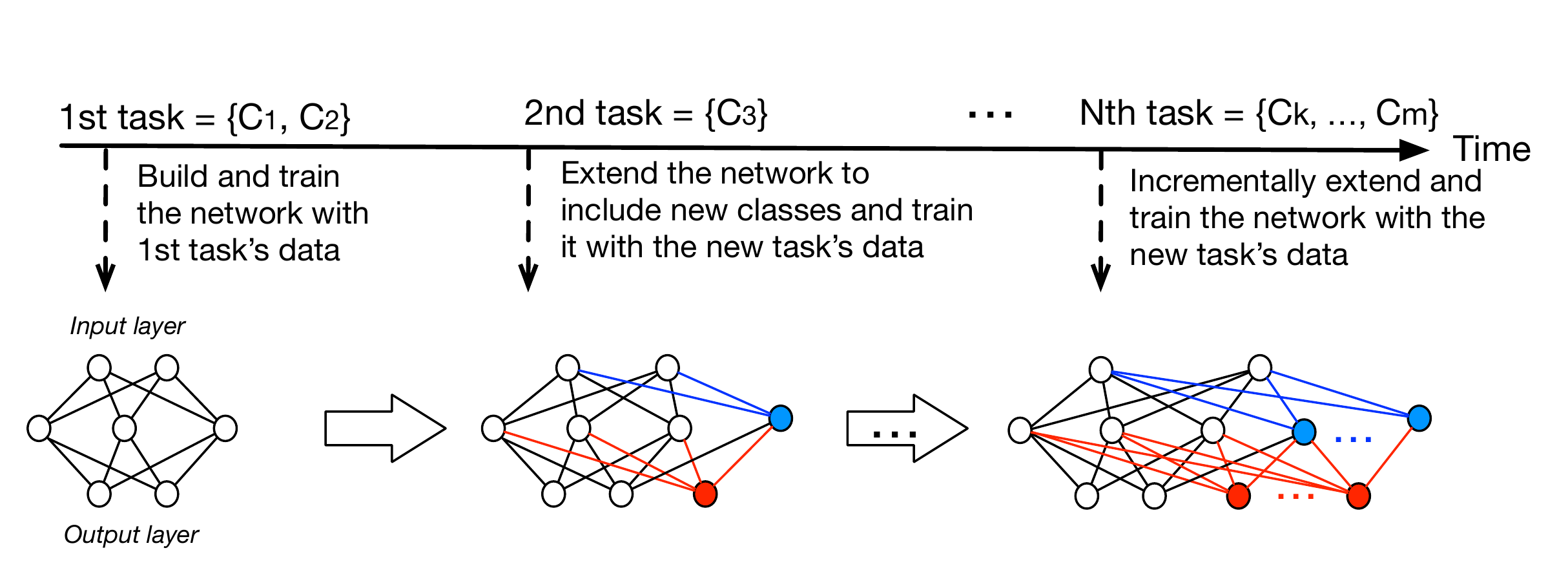}
\caption{An example of task-incremental continual learning with neural network}
\label{fig:CL_example}
\end{figure}

\begin{figure}[!htp]
\centering
\includegraphics[width=0.6\textwidth]{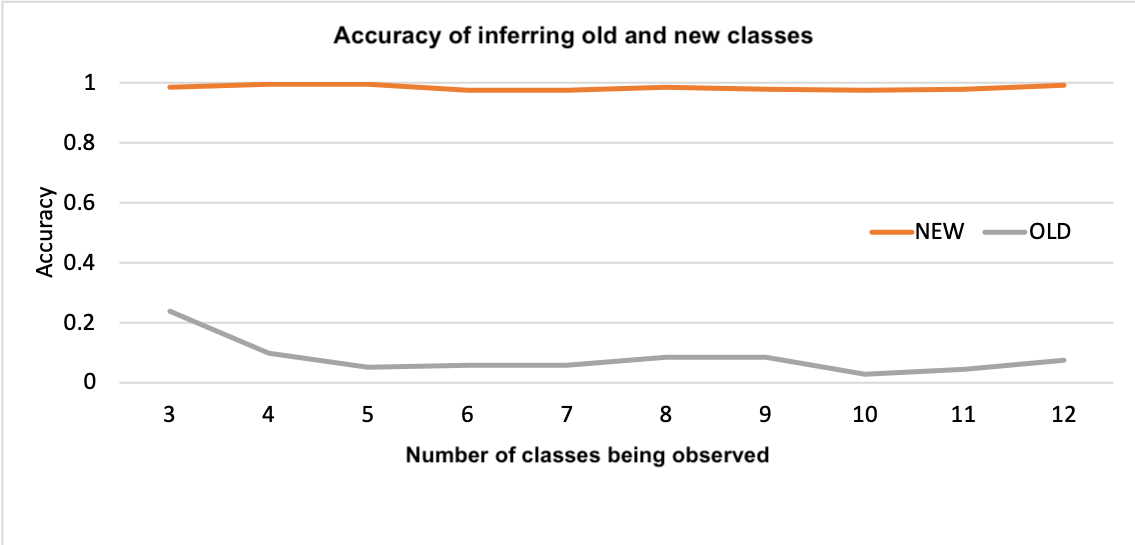}
\caption{An example of catastrophic forgetting effect~\cite{ye2020evolving} where the accuracy on new classes is high and the accuracy on old classes remains low as the network keeps forgetting the old knowledge}
\label{fig:CL_acc}
\end{figure}

Continual learning, or lifelong learning, is evoking increasing attention in the field of machine learning, aiming to retaining old knowledge and accumulating new knowledge by continually learning new tasks over time~\cite{IRFAN202180,Kemker2018b,farquhar2018robust,pflb2019comprehensive,lange2019continual,lesort2019continual}. In HAR, we are concerned with \textit{task-incremental} continual learning~\cite{10.3389/frai.2020.00019}, where a system learns one task at a time and each task contains training data on a new set of classes. Take a neural network as an example in Figure~\ref{fig:CL_example}, where for each new task, the network needs to be extended to include new classes and/or add more parameters, and then trained with the new task's data. Faced with this continual learning setting is often the \textit{catastrophic forgetting} (CF) problem; that is, the network will be optimised towards the new task's data while forgetting the old knowledge and thus degrade the performance on inferring the old classes. As presented in the example of Figure~\ref{fig:CL_acc}, after training on each new task, the network can only achieve high accuracy on the current classes, while obtaining low accuracy on all the old classes. 

A large number of continual learning techniques with neural networks have devoted to tackling this CF challenge~\cite{lange2019continual,PARISI201954}: (1) regularisation -- constraining the network parameter updates such that old knowledge is retained; (2) rehearsal -- storing a small number of old classes' data to replay with new classes' data when training a network; and (3) dynamic architectures -- extending the network architecture for new knowledge while keeping the parameters important for the old classes.

The purpose of this paper is to benchmark the state-of-the-art continual learning techniques on sensor-based human activity recognition datasets via extensive empirical evaluation\footnote{The code and all the experiments are available at \url{https://github.com/srvCodes/continual-learning-benchmark}.}. We exclusively focus on neural network based techniques as they have been widely adopted in HAR tasks~\cite{Chen2020DeepLF}. Continual learning techniques have achieved promising performance in computer vision~\cite{Kemker2018b,farquhar2018robust,pflb2019comprehensive} and robotics~\cite{lesort2019continual}, and it would benefit to assess to what degree they can tackle the continual learning challenges in HAR, as unlike image data, sensor data exhibits its limitations. 
\begin{itemize}[noitemsep,topsep=0pt,parsep=0pt,partopsep=0pt]
    \item \textit{Scarce labels} -- In \ac{HAR}, it is very time- and effort-consuming to label sensor data from real-world deployment. As continual learning is concerned with new, unobserved, unpredicted activities, it relies on users' self-annotation; e.g., a user provides a free-text label for the new activity that he/she is performing. Such labels can be sparse and noisy~\cite{LYU2021454}. Due to the labelling challenge, there is a smaller amount of labelled training data in HAR compared to their vision counterparts like CIFAR-100~\cite{Krizhevsky09learningmultiple}. 
    \item \textit{Imbalanced class distribution} -- Some activity classes can have a high occurrence frequency and dominate the dataset while others being rare. For example, hiking can be relatively infrequent compared to sleeping. 
    \item \textit{Sensor noise} -- Sensors can produce noisy readings due to unintended interactions with the environment or degradation over time~\cite{YE201632}. For example, a visitor or a pet can trigger unexpected sensor traces in a single-resident dwelling. 
\end{itemize} 

Also, sensor data in \ac{HAR} shares the characteristics with image data: (1) \textit{Intra-class diversity} -- one activity class can have multiple patterns due to different ways of performing an activity; \textit{e.g.}, different ways of preparing a meal. (2) \textit{Inter-class similarity} -- some activity classes have overlapping decision boundaries, which makes them difficult to separate. For example, reading and writing for a sedentary user can have very similar distributions in sensor feature space.

The above limitations of sensor data will bring extra complication in continual learning. This paper aims to explore to what extent the existing techniques can address continual learning challenges in \ac{HAR}. More specifically, the questions that we seek to answer are:
\textit{how computationally expensive are these in terms of memory and training time?},  \textit{is it affordable to run them on resource-constrained devices?}, or \textit{is their performance sensitive to the amount of training data?} To answer these questions, we adapt the existing continual learning evaluation methodology to HAR and design a general evaluation framework to assess 10 recent techniques published between 2016 and 2020 on 8 commonly used \ac{HAR} datasets including both ambient binary sensors and accelerometers. 
We focus on two types of continual learning techniques: regularisation and rehearsal based methods, which are most appropriate for HAR tasks. 

Our evaluation has uncovered the following findings. 
\begin{itemize}
    \item The rehearsal techniques significantly outperform regularisation techniques on our selected datasets, and the regularisation terms alone are not able to retain the old knowledge. The regularisation terms that tackle class imbalance and inter-class separation are most effective for HAR. 
    \item  Most of the rehearsal techniques do not need to store many samples in memory (e.g., 4 or 6 samples per class), and often random sampling can outperform the other sophisticated, computation-expensive techniques.
    \item The selected continual learning techniques are not sensitive to training data size, and training with 30\% of each dataset can achieve good accuracy.
    \item  The computation cost for most of the selected techniques is relatively low; \textit{i.e.}, around 33 seconds for training each incremental task. Therefore, these techniques are affordable on resource-constrained devices. 
\end{itemize} 

The paper is outlined as follows. Section~\ref{sec:problem_statement} defines the continual learning problem and briefly reviews the state of the art techniques in three categories:  regularisation, rehearsal, and dynamic architecture. Section~\ref{sec:techniques} presents a detailed description of the selected techniques. Section~\ref{sec:experiments} introduces the experimental setup including datasets, evaluation process, metrics and baseline. Section~\ref{sec:results} describes the results and Section~\ref{sec:discussion} discusses our findings to shed light on the future direction of continual learning in sensor-based \ac{HAR}. Finally, Section~\ref{sec:conclusion} concludes the paper.

%It is defined as the ability to preserve old knowledge in the model and learn new knowledge without retraining the model from scratch~\cite{}. Take a neural network as an example in Figure~\ref{}. The network is trained with training data on a set of classes. 

%A continual learning empowered network is initialised with training data on a set of classes. Then

% define continual learning in HAR

% why it is important

% why it is challenging 

% what we are trying to do in this paper and what's the highlight

\section{Problem Statement}\label{sec:problem_statement}
% Define lifelong learning problem in sensor based activity recognition
In this section, we define the setup for our continual learning problem in a task-incremental setting and briefly introduce the mainstream continual learning techniques. 

The task-incremental continual learning setting assumes tasks arriving in a sequential manner, where each task comprises one or more classes. Formally, let a task sequence $T$ of $N$ tasks be $ T = [t_1, t_2, t_3, .., t_N]$, and a task $t_i$ ($i \in [1, .., N]$) is coupled with a set of classes $C^i=\{c^i_1,..., c^i_{K_i}\}$ and a collection of training data $\{(\vec{x}^i_j, y^i_j)|y^i_j \in C^i\}_{j=1}^{M^i}$, where $K_i$ is the number of classes and $M_i$ is the number of training data in a task $t_i$. The learning objective on the $i$th task is to optimise a model to classify the current classes in $C^i$ and all the previous classes in $C^{i-1} \cup ... \cup C^1$, where $C^i \cap (C^{i-1} \cup ... \cup C^1) = \emptyset$; i.e., each task has a mutually exclusive set of classes and new classes in the current task have not been observed in the previous tasks.
% \todo[]{Martin: yes this is the class incremental setup. We should either say that we deal with the class incremental setup or generalise more}. 

% \todo[inline]{Shorten introduction and expand here. }

The major problem in task-incremental setting resides in \textit{catastrophic forgetting} (CF), where the new knowledge learnt by the model interferes with the old knowledge so that the performance on classifying old classes degrades over time. CF is caused by the stability-plasticity dilemma~\cite{ABRAHAM200573}. Plasticity refers to the ability of a model to accommodate new knowledge while stability refers to its ability to retain old knowledge. High plasticity often causes drastic changes in the model's parameters, thus interfering with the previous knowledge captured by them. Most of the existing continual learning techniques are trying to balance plasticity and stability, and these techniques are often grouped in three categories~\cite{PARISI201954}: (1) regularisation, (2) rehearsal, and (3) dynamic architectures.

\subsection{Regularisation Techniques}
Regularisation techniques often introduce additional terms to the loss function that constrains the weight updates of the network so as to prevent compromising the performance on old tasks. \ac{KD}~\cite{Hinton2015}, a way to transfer knowledge between different networks, has been widely adopted to retain the knowledge of old tasks when learning new ones. It prevents the current model's output deviating from the previous model's prediction that is recorded as logit output as soft labels for old classes. Learning without forgetting (LwF)~\cite{Li2016} is the earliest attempt to employ the distillation loss in continual learning, followed by many others such as cross-distilled loss~\cite{10.1007/978-3-030-01258-8_15} and attention distillation loss~\cite{dhar2018learning}. 

Another category of regularisation techniques is to identify important parameters for old tasks and penalises updates on them. For example, Elastic weight consolidation (EWC) applies Fisher Information matrix to measure the importance of the network parameters~\cite{Kirkpatrick2016}.  Memory Aware Synapses (MAS) estimates the importance of a parameter based on the magnitude of the gradient; that is, how much change the output of the network is caused by a small perturbation to the parameter~\cite{Aljundi2018}. 

Group sparsity regularisation has been applied to allow for selective training a subset of neurons. Adaptive Group Sparsity based Continual learning (AGS-CL)~\cite{Jung2020AGS} introduces regularisation terms using two node-wise group sparsity based penalties. The first term assigns and learns new important nodes via the ordinary group Lasso penalty when learning a new task, while the second term applies the group-sparsity based deviation penalty to prevents the drift on important node parameters. 

Specific regularisation terms have been designed to tackle specific problems. Weight alignment has been employed to balance the distribution of new tasks' training samples and old tasks' in-memory samples~\cite{zhao2019maintaining, Kim2019AdjustingDB}. Customised regularisation terms have been designed to prevent task interference; for example, forcing a large margin between the old and new classes~\cite{Hou2019}. 

%This type of techniques can mitigate catastrophic forgetting to a certain degree, and it often works better when used together with replay samples.

\subsection{Rehearsal Techniques}
Rehearsal techniques mitigate catastrophic forgetting by mixing data from previous tasks with the current task. The previous task data can be used as inputs for re-training the network or as a constraint to the network updates for penalising the interference with the previous tasks. iCaRL, a class-incremental learning technique, stores a small set of representative samples for each class in memory, and combines these samples with new task data to update the network every time~\cite{Rebuffi2017}. This type of techniques often requires extra memory space for old task samples and also can be prone to overfitting the stored data, instead of generalising to old tasks~\cite{lange2019continual}. REMIND utilises data compression and augmentation to enable more effective replay~\cite{REMIND}. 

In contrast, gradient episodic memory (GEM) uses these old task samples to impose a constraint to allow positive backward transfer; that is, preventing the gradient update from increasing the loss on previous tasks~\cite{Lopez-Paz2017b}. Similarly, orthogonal gradient descent (OGD)  maintains a space consisting of the gradient directions of the network predictions on previous tasks and projects the loss gradients of new samples perpendicular to this gradient before backpropagation~\cite{farajtabar2019orthogonal}. In this way, OGD minimises the interference on old tasks and thus preserves old knowledge. 

In addition to sampling from training data, in-memory samples on previous tasks can also be generated via a generative model that learns the distribution of old tasks~\cite{DGR,Rios_Itti18nipscl,Ostapenko_Puscas2019,conf/iccv/ZhaiCTHNM19}. Generative replay model (GRM)~\cite{DGR} employs a Generative Adversarial Network (GAN) for generating samples on previous tasks for \textit{pseudo rehearsal}. This type of techniques relies on the quality of generated samples; for example, GANs might suffer mode collapse in that the generated samples are clustered in one specific space, rather than diverse across the whole space. Also training GANs can add extra computation cost to training and GANs' performance can degrade over time with more and more classes being learnt~\cite{lange2019continual}. 

\subsection{Dynamic Architectures and Ensemble Techniques}
Dynamic architectures tackle catastrophic forgetting by retaining the model on the old tasks while extending it with new parameters to learn new tasks. They are closely tied with ensemble methods which train multiple models for different tasks.
\textit{ExpertGate}~\cite{expertgate} consists of a network of experts where each \textit{expert} is a model trained on a specific task. A gating mechanism decides which expert is required for activation. This bypasses the need for loading all models, which is memory efficient as each model can be computationally intensive~\cite{chen_liu_2018}. Net2Net~\cite{net2net} is another type of dynamically evolving network which can be widened (adding more neurons) and deepened (adding more layers). Knowledge from the previous network is preserved in the newly constructed network.

\textit{Progressive network}~\cite{prognet} keeps a pool of models that are pre-trained with previous knowledge and adds lateral connections to them for a new task \cite{chen_liu_2018}. To mitigate forgetting, the parameters for previous tasks are never modified while new parameters are learned for the new task. Therefore, it does not deteriorate the performance of previous tasks. One drawback of using this technique is that the network can become complex with an increasing number of tasks learned. Since a new network is learned for each task and it needs to be connected to the previous network, the complexity of the network structure and parameters increases quickly~\cite{chen_liu_2018}.  

Dynamically Expandable Networks (DEN)~\cite{yoon2018lifelong} allows layer expansion and employs group sparsity regularisation to identify the neurons that are relevant to new tasks and allow selective retraining on these neurons. Compacting, Picking, and Growing (CPG) continual learning~\cite{hung2019compacting} adds new neurons for accommodating new tasks and then constantly applies network compression by deleting unnecessary weights. 
 
Dynamic architectures often have been employed in computer vision applications, where there exist hundreds or thousands of classes. As our selected \ac{HAR} datasets do not have such a high number of activities to learn and a decent size of a neural network often works well, we exclude this category of techniques in our study.

% A wide range of continual learning techniques in computer vision and robotics have been covered in recent surveys~\cite{PARISI201954,lesort2019continual} with qualitative analysis and comparison. De Lange et al.~\cite{lange2019continual} have designed comprehensive evaluation methodology and performed empirical evaluation on a set of representative techniques. Other surveys quantitatively assess a smaller set of techniques with a particular focus on measuring catastrophic forgetting~\cite{Kemker2018b,farquhar2018robust,pflb2019comprehensive}. In this paper, we select a small collection of representative techniques and empirically assess their performance on a variety of HAR datasets. 

\section{Comparison of Techniques}\label{sec:techniques}
After presenting the overview of continual learning techniques in the previous section, now we will focus on a small set of techniques. We set three criteria for selecting state-of-the-art continual learning techniques: (1) the techniques are built on neural networks and target classification tasks, as neural networks have been widely adopted in HAR tasks for their effectiveness at feature extraction and recognition~\cite{WANG20193}, 
(2) the techniques should be the most representative ones, which have been included in several recent continual learning surveys~\cite{lange2019continual,Kemker2018b,PARISI201954}, and (3) the techniques target HAR limitations including class imbalance and inter-class separation. With these criteria, we select 10 techniques from regularisation and rehearsal categories and in the following, we will provide a brief description of each and illustrate their key characteristics.

\subsection{Regularisation Techniques}
\noindent\textbf{Learning without Forgetting}~\cite{Li2018} aims to keep the output of the old tasks from the new network close to the output from the original model. This is achieved by using the knowledge distillation \ac{KD} loss as a regularisation term. \ac{KD} is a way to transfer knowledge from a complex \textit{teacher} model to a simpler \textit{student} model by minimising the loss on the output class probabilities from the teacher model~\cite{Hinton2015}. \ac{KD} has been widely applied to various continual learning techniques~\cite{10.1007/978-3-030-01258-8_15,dhar2018learning, Li2018} to distil knowledge learnt from old tasks to the model for new tasks, which is defined as follows: 
\begin{align}\label{eq:kd_loss}
     \mathcal{L}_{KD}(y_o, \hat{y}_o) = -\sum\limits_{i=1}^{l}y_o^{'(i)}\log\hat{y}_o^{'(i)}, 
\end{align}
where $l$ is the number of class labels, and $y_o^{'(i)}$ and $\hat{y}_o^{'(i)}$ are temperature-scaled \textit{recorded} and \textit{predicted} probabilities of the current sample for an old class label $i$. The temperature is used to tackle the over-confident probability that the teacher model produces on their prediction. That is, 
\begin{align*}
y'^{(i)}_o = \frac{(y^{(i)}_o)^{1/T}}{\sum_j(y^{(j)}_o)^{1/T}} \text{ and } \hat{y}'^{(i)}_o = \frac{(\hat{y}^{(i)}_o)^{1/T}}{\sum_j(\hat{y}^{(j)}_o)^{1/T}}.
\end{align*} 
The loss $\mathcal{L}_{KD}$ is combined with the cross-entropy loss on a new task's samples $\mathcal{L}^n_{CE}$; that is, 
\begin{align}
&\mathcal{L}(y_n,\hat{y}_n,y_o, \hat{y}_o) = \lambda_{o}\mathcal{L}_{KD}(y_o, \hat{y}_o) + \mathcal{L}_{CE}(y_n, \hat{y}_n), \\
&\mathcal{L}^n_{CE}(y_n, \hat{y}_n) = -y_n \log \hat{y}_n,    
\end{align}
% \todo[inline]{Martin: original lwf paper does not use balace weight?}
where $y_n$ is the one-hot encoded vector of the ground-truth label, $\hat{y}_n$ is the predicted logit (i.e., the softmax output of the network) on new class labels, and $\lambda_o$ is a loss balance weight. A larger $\lambda_o$ will favour the old task performance over the new task. During training, a new batch gets first fed through the old network to record its outputs, which then are used for $\mathcal{L}_{KD}$ so that the network updates will not deviate from the old network.

\noindent\textbf{Elastic Weight Consolidation} (EWC)~\cite{Kirkpatrick2016} draws inspiration from research on mammalian brains, which suggests that catastrophic forgetting can be avoided through the protection of synapses. EWC tries to mimic this by inhibiting changes on parameters that are deemed important for previous tasks. The importance of parameters is modelled as the posterior distribution $p(\theta | D)$; that is, optimising the parameters is to find their most probable values with respect to some data $D$. In the context of continual learning, if the data $D$ is assumed to consist of two independent tasks $A$ and $B$, then 
\begin{align*}
\log p(\theta|D) = \log p(D_B|\theta) + \log p(\theta|D_A) - \log p(D_B).
\end{align*}
The posterior probability $p(\theta|D_A)$ contains information about which parameters are important to task $A$. However, the true probability $p(\theta|D_A)$ is intractable, and thus it is estimated via Laplace approximation~\cite{MacKay1992} as a Gaussian with diagonal precision determined by the Fisher Information Matrix (FIM). The loss function of EWC is defined as: 
\begin{equation}\label{eq:ewc}
   \mathcal{L}(\theta) = \mathcal{L}^B_{CE}(\theta)+\sum\limits_i\frac{\lambda}{2}F_i(\theta_i-\theta^*_{A,i})^2, 
\end{equation}
where $\mathcal{L}^B_{CE}$ is the loss on the new task B only, $\lambda$ indicates the importance of the old task A with respect to B, and $i$ is the parameter index. Originally, one FIM is required for each task, and later it can be resolved by propagating them into a quadratic penalty term~\cite{Huszr2018NoteOT}. However, this formulation assumes the FIM to be diagonal, which is not always the case. Rotating EWC (R-EWC)~\cite{xialei2018forgetting} improves upon EWC by reparameterising $\theta$ through rotation in a way that it does not change outputs of the forward pass but the computed FIM is approximately diagonal. 

\noindent\textbf{Memory Aware Synapse} (MAS)~\cite{Aljundi2018}, inspired by neuroplasticity, also considers the importance of network parameters, which is measured in an online, unsupervised manner. Here, the importance is approximated by the sensitivity of the learned function to a parameter change. Given a data point $x_k$, the network output is defined as $F(x_k; \theta)$. A change in the network output caused by a small perturbation $\delta = \{\delta_{ij}\}$ in the parameters $\theta=\{\theta_{ij}\}$ can be approximated as: 
\begin{align*}
F(x_k;\theta+\delta) - F(x_k;\theta) \approx \sum\limits_{i,j}g_{ij}(x_k)\delta_{ij}\\
g_{ij}(x_k) = \frac{\partial(F(x_k,\theta)}{\partial\theta_{ij}}.
\end{align*}
where $g$ is the gradient with respect to the parameter $\theta$. By accumulating gradients over all the data points, the importance weight on a parameter $\theta_{ij}$ is computed as: 
\begin{align*}
\Omega_{ij} = \frac{1}{N}\sum\limits_{k=1}^N||g_{ij}(x_k)||.
\end{align*}
When learning a new task, the loss function is defined as:
\begin{equation}\label{eq:mas}
    \mathcal{L}(\theta) = \mathcal{L}^n_{CE}(\theta) + \frac{\lambda}{2}\sum\limits_{i,j}\Omega_{ij}(\theta_{ij}-\theta^*_{ij})^2, 
\end{equation}
where $\mathcal{L}^n_{CE}(\theta)$ is the CE loss on the new task, $\theta_{ij}$ and $\theta^*_{ij}$ are the new and old network parameters, and $\lambda$ is a hyperparameter that indicates the regularisation strength. The idea is to retain the old knowledge by penalising large updates on the sensitive network parameters. 

\subsection{Rehearsal Technique}
\noindent\textbf{Incremental Classifier and Representation Learning} (iCaRL)~\cite{Rebuffi2017} is an early attempt of rehearsal approaches for class-incremental learning. It leverages memory replay and regularisation. After training on each task, it employs the herding sampling technique~\cite{Welling2009} to select a small number of exemplars (i.e., representative samples) from the current task's training data and stores them in memory. Herding works by choosing samples that are closest to the centroid of each old class. When the next task becomes available, the in-memory exemplars will be combined with the new task's training data to update the network. The loss function of iCaRL is the same as Equation~\ref{eq:kd_loss} in LwF, which is a combination of CE loss on new classes and the KD loss on old classes to allow knowledge transfer. 

% minimises the CE loss on new classes and the KD loss of old classes between previous and current network outputs. The classification is done through the nearest-of-mean classifier.

\noindent\textbf{Incremental Learning In Online Scenario} (ILOS)~\cite{He2020IncrementalLI}, similar to iCaRL, also employs memory replay with KD loss for regularisation. The key difference is that ILOS  uses an updated version of the CE loss. It introduces an accommodation ratio $0 \leq \beta \leq 1$ to adjust the proportion of logits between the current and  previous model:
\begin{equation}\label{eq:ilos-y}
    \Tilde{y}'^{(i)} = \begin{cases}
   \beta y^{(i)} + (1 - \beta)\hat{y}'^{(i)}, &  1 \leq i \leq n \\
   y^{(i)},  & n+1 \leq i \leq n+m
\end{cases}
\end{equation}
where the indices $[1,n]$ refer to old classes, $[n+1, n+m]$ refer to new classes, $y^{(i)}$ are the one-hot encoded output logits of the current model and $\hat{y}'^{(i)}$ are the recorded old classes' output logits from the previous model. In this way, the output from the previous model will be retained in the current network and the retaining degree is regulated via the parameter $\beta$. The larger the $\beta$, the more retained the output from the previous model. While the KD loss is still calculated using $y^{(i)}$, the CE loss is now based on the adjusted output $\Tilde{y}'^{(i)}$ instead of the recorded output $\hat{y}^{(i)}$:
\begin{equation}\label{eq:ilos-ce}
    \Tilde{\mathcal{L}}_{CE} = \sum^{n+m}_{i=1} - y^{(i)}\log[\Tilde{y}'^{(i)}].
\end{equation}
The final loss is the combination of the above CE loss and KD loss: 
\begin{equation}\label{eq:ilos}
    \mathcal{L}_{ILOS} = \alpha L_{KD} + (1-\alpha) \Tilde{\mathcal{L}}_{CE}.
\end{equation}
Following the guideline in the original paper, both $\alpha$ and $\beta$ can be set to 0.5.

\noindent\textbf{Gradient Episodic Memory} (GEM)~\cite{Lopez-Paz2017b} alleviates forgetting by controlling the gradient updates to balance the performance on old and new classes. GEM uses a memory space to host examples from previous classes $\cup_{k \in [1, t-1]} M_k$, where $M_k$ is the samples stored on a previous task $k$ ($k \le t$) and $t$ is the current task's index. When training a new task $t$, $M_k$ is used in an inequality constraint to avoid an increase in loss. Hence, the loss of a current task $\mathcal{L}$ must be smaller than or equal to the loss of the previous model: 
\begin{equation}
    minimize_{\theta} \quad \mathcal{L}(f_{\theta},D_t) \quad s.t. \quad \mathcal{L}(f_{\theta},M_{k})\leq \mathcal{L}(f_{\theta}^{t-1},M_k)\qquad \forall k<t
\end{equation}
where $f_\theta$ represents the model with parameters $\theta$. To achieve this, GEM restricts the angle between gradient vectors $g_k$ of previous and current task $g_t$ to be no greater than 90\degree. If the angle is greater than 90\degree, the new gradient vector gets projected to the euclidean distance closest vector $\tilde{g}$ that is inside the allowed range. Thus, the optimisation problem can be defined as:
\begin{equation*}
    minimize_{\tilde{g}} \quad \frac{1}{2}||g-\tilde{g}||^2_2 \quad s.t. \quad \langle\tilde{g},g_k\rangle\geq0\qquad \forall k<t
\end{equation*}
Solving this quadratic program problem for all in-memory samples is very computationally expensive. To address the computational cost, A-GEM~\cite{Chaudhry2019} is proposed to ensure that there is no increase in the average loss over the episodic memory. That is, A-GEM only uses a smaller set of in-memory samples to calculate an average gradient instead of all gradients for a task. 
% In our experiments, A-GEM does not outperform GEM so we use the original GEM.
% \todo[inline]{Martin: no comment about us not using a-gem?}
% Thus, reducing the complexity of the optimisation problem to a single constraint resulting in the following objective and optimization problem:
% \begin{align*}
%     &minimize_{\theta} \quad \mathcal{L}(f_{\theta},D_t) \quad s.t. \quad \mathcal{L}(f_{\theta}, M\leq \mathcal{L}(f_{\theta}^{t-1},M)\qquad \text{ where } M=\cup_{k<t}M_k\\
%     &minimize_{\tilde{g}} \quad \frac{1}{2}||g-\tilde{g}||^2_2 \quad s.t. \quad \tilde{g}^T,g_{ref}\geq0.
% \end{align*}

\noindent\textbf{Learning a Unified Classifier Incrementally via Rebalancing} (LUCIR)~\cite{Hou2019} was proposed to tackle the imbalance between a small number of in-memory samples from old tasks and a larger number of samples from a new task. The imbalance results in the training being biased towards new tasks. LUCIR proposes multiple (loss-) components to mitigate this adverse effect. Firstly, cosine normalisation is used in the last layer to level the difference of the embeddings and biases between all classes since those are significantly higher for new tasks. Secondly, less-forget constraint is introduced to prevent forgetting old classes' geometric configurations by encouraging the extracted features of new classes similarly rotated to those of old ones; that is: 
% $\mathcal{L}_{dis}^G(x)= - \sum_{i=1}^{|C_0|} || \langle\tilde{\theta}_i}, \tilde{f}(x)\rangle - \langle\tilde{\theta}^*_i}, \tilde{f}^*(x)\rangle ||$, where $f^*$ and $\theta^*$ are the features and network outputs in the original model, and $|C_0|$ is the number of old classes. 
\begin{equation}\label{eq:lessforget}
 \mathcal{L}_{dis}^G(x)= 1 - \langle\Tilde{f}^*(x),\Tilde{f}(x)\rangle,   
\end{equation}
where $\Tilde{f}(x)$ and $\Tilde{f}^*(x)$ are normalised features generated by the new and old model, and $\langle, \rangle$ measures the cosine similarity between the two normalised vectors. Thirdly, a margin ranking loss is used to enhance inter-class separation. For each in-memory sample $x$, it aims to separate old classes from all the new classes by a margin. Using $x$ as an anchor, LUCIR finds positive and negative embeddings and aims to maximise their distances. The positive embeddings are from the ground-truth class of $x$ (represented as $\Tilde{\theta}(x)$), while the negative embeddings are from the top-K new classes that produce the highest response to $x$ (represented as $\Tilde{\theta}^k$), indicating the classes that $x$ is mostly confused with. Then the margin ranking loss is defined as: 
 \begin{equation}\label{eq:lucir-mr}
 \mathcal{L}_{mr}(x)=\sum\limits_{k=1}^K\max(m-\langle\Tilde{\theta}(x),\Tilde{f}(x)\rangle+\langle\Tilde{\theta}^k.\Tilde{f}(x)\rangle,0).
 \end{equation}
This leads to the  combined loss function as follows:
\begin{equation}\label{eq:LUCIRcomb}
    L=\frac{1}{|N|}\sum\limits_{x\in N}(L_{CE}(x)+\lambda L_{DIS}^G(x))+\frac{1}{|N_o|}\sum\limits_{x\in N_o}L_{MR}(x), 
\end{equation}
where $N$ refers to all the training samples and $N_o (\subset N)$ refers to the reserved old samples. $\lambda$ is a dynamic weight to adjust how much knowledge of the previous model needs to be preserved depending on how many new classes are introduced. It is calculated by multiplying the base $\lambda$ with the squared root of the ratio between new and old classes; that is, $\lambda=\lambda_{base}\sqrt{|C_N|/|C_o|}$. It regulates the degree of preserving the old knowledge by taking into account the number of new classes being added. For example, when there are many new classes are introduced, the model would preserve less old knowledge to allow the model to adapt to the new knowledge, and vice versa. In general,  $\lambda_{base}$ is set as 5.0 and the margin value $m$ in Eq~\ref{eq:lucir-mr} as 0.5~\cite{Hou2019}.

\noindent\textbf{Weight Alignment for Maintaining Discrimination and Fairness} (WA-MDF) is another approach that tackles the mentioned imbalance problem by correcting the biased weights in the last fully connected ({FC}) layer after training each task. It aligns the norms of the weight vectors of new classes to those of old classes. The {FC} layer output of a model can be expressed in a general form: $o(x) = W^T\phi(x)$, where $\phi(x)$ is the feature extraction function of an input $x$ and $W$ is the weight vector of the {FC} layer. $W$ can be separated into $W = (W_{o}, W_{n})$, where $W_{o}$ and $W_{n}$ refers to the weights corresponding to the old and new classes; that is, 
\begin{align*}
&W_o=(w_1, w_2, ..., w_{C^b_{o}}) \\
&W_n = (w_{C^b_{o}+1}, ..., w_{C^b_{o}+C^b}),
\end{align*}
where $C^b$ is the total number of classes and $C^b_{o}$ is the number of old classes. Their norms are defined as: 
\begin{align*}
&Norm_{o} = (||w_1||, ||w_2||, ..., ||w_{C^b_{o}}||) \\
&Norm_n = (||w_{C^b_{o}+1}||, ..., ||w_{C^b_{o}+C^b}||).
\end{align*}
The normalised weights on new classes are $\widehat{W}_n = \gamma W_n$, where $\gamma = \frac{Mean(Norm_o)}{Mean(Norm_n)}$. In this way, the average norm of the weights on the new classes will be the same as that on the old classes. The corrected output logits from the {FC} layer will be written as: 
\begin{equation}\label{wa1_eqn}
        O_{corrected}(x) = (W^T_{o} \phi(x), \gamma W_{n}^T \phi(x))^T.
    \end{equation}

As a norm-control method; {i.e.}, keeping a check on the norms of class embeddings following each gradient update, WA-MDF clamps the parameters of the {FC} layer at zero. Assuming that a network usually employs variants of ReLU activation units, the clipping facilitates the projection of weights. This helps eliminate large negative elements of weight vectors thus making its norm more consistent with the non-negative output logits of the ReLU function. While such a restriction seems to interfere with the convergence of the model, several studies have shown that training neural networks with weights projected via such distortions makes them robust to other types of distortions~\cite{Merolla2016DeepNN, Courbariaux2015BinaryConnectTD}. Here, distortion refers to different ways of weight projection operations; for example, 
adding Gaussian noise or performing additive, multiplicative, or power operations on weights~\cite{Merolla2016DeepNN}. Such distortion allows weights to accumulate small gradient updates and thus leads to improved performance. 

\noindent\textbf{Weight Alignment for Adjusting Decision Boundary} (WA-ADB) is also proposed to re-scale the weight vectors of majority and minority classes, but in a more general class-imbalance scenario~\cite{Kim2019AdjustingDB}. Different from the above WA-MDF, WA-ADB re-scales the weights based on the sample ratio of old and new classes instead of their weight norms. For a dataset \textit{D} with \textit{K} classes, given that $n_i$ is the number of samples of class \textit{i} ($n_1 \geq ... n_i ... \geq n_K$), the weight vectors of each class are re-scaled by an exponent of the re-scaling factor $n_1 / n_i$; {i.e.},
$w_i = (\frac{n_1}{n_i})^\gamma . w_i$. During training, the weight vectors will be normalised after each gradient update. While a larger $\gamma$ scales up the volume of feature space of a model allocated to infrequent classes, weight vector normalisation forces the class conditional probability to have the same variance independent of the sample size. 

To adapt WA-ADB to incremental learning scenario, we make the following adjustment. For an incremental training step \textit{i}, the dataset $D$ is the combination of in-memory samples on old tasks and new training samples on a new task. We then re-scale the weights of all the classes of old tasks by the same factor, i.e., $n_1 / n_i$, where $n_1$ is the number of samples on the largest class that has been seen so far and $n_i$ is the holdout size. Because $n_i$ is fixed due to the memory size, we only compute this number once. While the bias term is dropped in the original work~\cite{Kim2019AdjustingDB}, we take into account that the biases for classes of new tasks also have norms larger in average than that of classes from old tasks. As a result, we re-scale the biases of each class by the corresponding weight re-scaling factor computed in the above step.

%  \begin{figure*}[!htbp]
%     \centering
%     \includegraphics[width=0.6\textwidth]{Figures/norm_vis_scatter_bias_only.png}
%     \caption{Norms of biases after training over two incremental tasks with four classes each on the PAMAP2 dataset, {\footnotesize showing that the biases are larger in average for classes from new task than those from older tasks.}}
%     \label{fig:norms_by_batches}
% \end{figure*}

\noindent\textbf{Bias Correction} (BiC)~\cite{8954008} introduces a BiC correction layer after the last {FC} layer to adjust the weights in order to tackle the imbalance problem. There are two stages of training. In the first stage, the network will be trained with new task data and in-memory samples of old tasks using the CE and KD loss similar to iCaRL. In the second stage, the layers of the network are frozen and a linear BiC layer is added to the end of the network and trained with a small validation set consisting of samples from both old and new tasks. The linear model of the BiC layer corrects the bias on the output logits for the new classes: 
\begin{equation*}
    \Tilde{y}_k = \begin{cases}
   \hat{y}_k&  1 \leq i \leq n \\
    \alpha \hat{y}_k + \beta,  & n+1 \leq i \leq n+m
\end{cases}
\end{equation*}
where $\hat{y}_k$ is the output logits on the $k$th class, the old classes are $[1,..., n]$ and the new classes are $[n+1, ..., n+m]$.  $\alpha$ and $\beta$ are the bias parameters of the linear model, which are optimised in the following loss function: 
\begin{align*}
L_{BiC} = -\sum_{k=1}^{n+m} \delta_{y=k} \log \mathtt{softmax}(\Tilde{y}_k),
\end{align*} 
where $\delta_{y=k}$ is the indicator function to check if the ground-truth label $y$ is the same as a class label $k$. The intuition is that a balanced small validation set for all seen classes can counter the accumulated bias during the training of the new task. As the non-linearities in the data are already captured by the model's parameters, the linear function proves to be quite effective in estimating the bias in the output logits of the model arising. 
% due to the class imbalance in train data.

% With the arrival of incremental batches from new tasks, the \ac{FC} layer begins to accumulate a bias towards new classes due to the growing imbalance in the train data. A class-balanced validation data for all seen classes thus helps counter this bias while training the classifier for the next incremental batch. Further, since the non-linearities in the data are already captured by the model's parameters, the linear function proves to be quite effective in estimating the bias in the output logits of the model arising due to the class imbalance in train data. 

So far we have described all the 10 selected continual learning techniques and in the next section, we will introduce the evaluation framework and methodology for assessing and comparing them on the HAR datasets.

\section{Experimental Setup and Evaluation Methodology}\label{sec:experiments}
The main objective of this paper is to assess \textit{to what degree} the state-of-the-art continual learning techniques can enable continual activity recognition. More specifically, we will seek to answer the following questions:
\begin{enumerate}
    \item What is the performance of these continual learning techniques on \ac{HAR} datasets?
    \item Which techniques best balance plasticity and stability when incrementally learning new activities?
    \item Which regularisation term works best for what type of data? 
    \item What is the impact of in-memory sample sizes on the accuracy of these techniques?
    \item Are these techniques sensitive to the size of training data?
    \item What is the computation cost of these techniques in terms of memory and training time? 
\end{enumerate}

To answer these questions, we have selected 8 HAR datasets (in Section~\ref{subsec:datasets}), design the evaluation process (in Section~\ref{subsec:process}), select the evaluation metrics (in Section~\ref{subsec:metrics}) that are appropriate for HAR, describe the baseline approach for comparison (in Section~\ref{subsec:baseline}), and illustrate the architecture configuration and hyperparameter tuning for all the techniques (in Section~\ref{subsec:architecture}). 

% \todo[inline]{Martin: I think the whole section needs some reformatting because right now there is no structure that says what we are doing where we can easily add the different steps and analysis we have performed. E.g. we do this to see the effect of... We also do that to...}

\subsection{Datasets}\label{subsec:datasets}
\begin{figure}[!htbp]
    \centering
    \includegraphics[width=\textwidth]{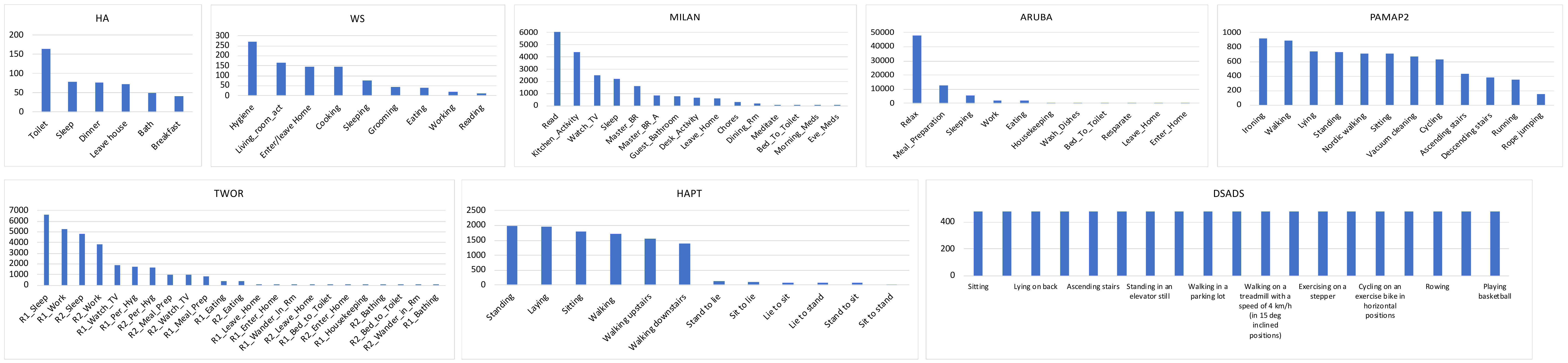}
    \caption{Activity histograms on all the selected datasets.}
    \label{fig:data-hist}
\end{figure}

To present a comprehensive profile of selected continual learning techniques, we will assess on a wide variety of most representative, state-of-the-art HAR datasets, ranging from simple datasets (with a single user, a small number of features and activity classes) to more complex datasets (with multiple users, a large number of features and activity classes). Driven by these criteria, we select the following set of publicly available, third-party datasets that are collected on accelerometer and event-driven binary sensor data, as these two are the most common sensor types in HAR. Table~\ref{tab:datasets} and Figure~\ref{fig:data-hist} summarise the main characteristics of these datasets. 
\begin{table}[!htbp]
    \centering
    \caption{Main characteristics of selected datasets}
    \label{tab:datasets}
    \includegraphics[width=0.8\textwidth]{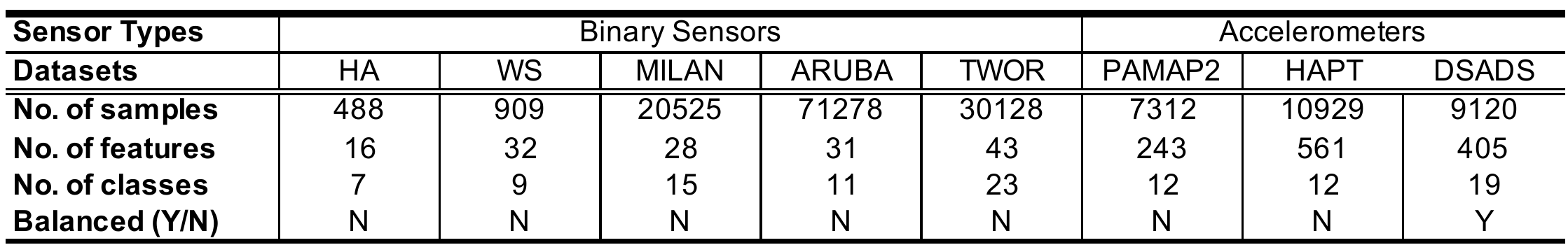}
\end{table}

\noindent\textbf{Binary sensor data} We use House A (denoted as \textit{HA}) from the University of Amsterdam~\cite{vanKasteren2011} and 4 CASAS smart home datasets (denoted as \textit{WS}, \textit{Milan}, \textit{Twor}, and \textit{Aruba}) from Washington State University~\cite{cook09dataset}. They are collected on a wide range of event-driven binary sensors, including passive infra-red sensors, state-change sensors, switch sensors, pressure sensors, and water flow sensors. On these datasets, we apply state-of-the-art techniques~\cite{8667728} to segment the raw sensor data into a 60-second interval and generate features as an activation ratio of each sensor in the interval.

\noindent\textbf{Accelerometer sensor data} 
Physical Activity Monitoring Dataset (\textit{PAMAP2})~\cite{6246152} contains 12 activities such as sitting, lying, and house cleaning. The data is collected on 9 users with 3 accelerometer units on each user's chest, dominant arm and side ankle. Daily and Sports Activities Dataset (\textit{DSADS})~\cite{ALTUN20103605,Altun:2010:HAR:1881331.1881338} contains 19 activities such as running, rowing, and sitting. The data is collected on 8 users with 5 accelerometer units on each user's torso, arms, and legs. On these datasets, we use the sensor features that have been extracted from raw accelerometer data; that is, on each accelerometer unit, 27 features are generated, including mean, standard deviation, correlations, and spectrum peak position. Human Activity Recognition Dataset (\textit{HAPT})~\cite{REYESORTIZ2016754} contains 12 daily activities collected on 30 subjects with a smartphone (Samsung Galaxy S II) on their waist. There are 561 features extracted from accelerometer and gyroscope readings. 

% Here we use simple feature extraction techniques on both binary and accelerometer sensor data and we believe that more sophisticated feature extraction techniques such Convolutional Neural Network and Recurrent Neural Network~\cite{WANG20193} will lead to better recognition accuracy. We consider feature extraction is out of scope of this paper. 

\subsection{Evaluation Process}\label{subsec:process}
% \todo[inline]{Martin: three-scenarios paper here: justify and explain the evaluation process and why our results are worse.}
% Different continual learning scenarios have been tested in computer vision~\cite{ven2019scenarios}, and here we focus on a more realistic, harsher setting that best resembles HAR problem - class-incremental learning with a single-output head. All the selected techniques produce much lower accuracy on this setting, compared to other set-ups. For example, permutation MNIST is often assessed~\cite{Kemker2018b}, where all the tasks are classifying the 10 digits and the only difference is that each task applies different permutations to the MNIST image pixels. Besides, multi-output-heads are often used, where the network knows \textit{a priori} which task it currently faces. These settings are 
% regarded inappropriate for \ac{HAR}.
We follow  the state-of-the-art task-incremental evaluation methodology~\cite{ven2019scenarios}, where we assign 2 randomly sampled classes to each task and form a sequence of $|C|/2$ consecutive tasks, where $|C|$ is the total number of classes in a dataset. In practice, the number of classes in each task can vary, depending on the arrival of new classes and update cycle of the model. Here, we set the number of classes as 2, as a good trade-off to assess the incremental learning capability and the training time. 

Also, we employ the stratified train-test split~\cite{ven2019scenarios} to split each class's data  into 70\% for training and 30\% for testing. As accelerometer datasets often have multiple users, to avoid data leakage, we split training and testing data of each class by users; that is, we use 70\% of users' data for training and the remaining 30\% of users' data for testing. 

\begin{figure}
    \centering
    \includegraphics[width=0.8\textwidth]{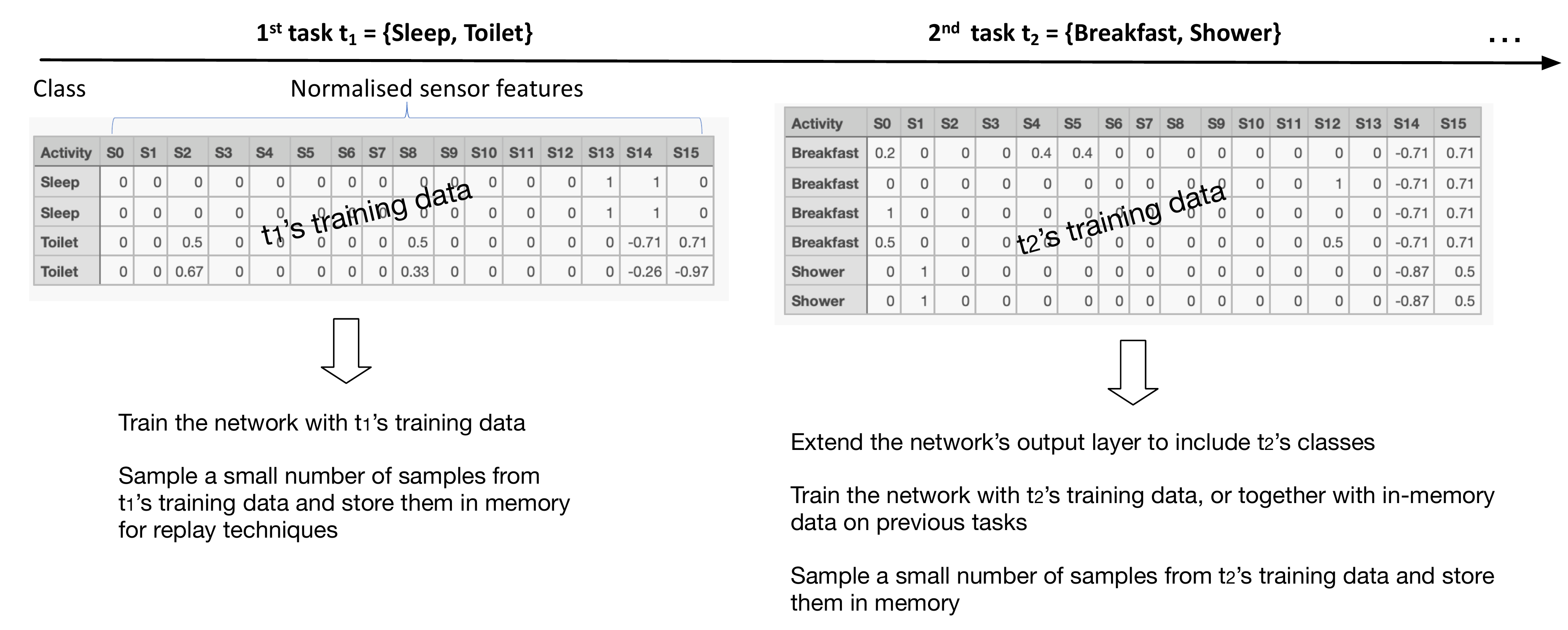}
    \caption{An example of task-incremental evaluation}
    \label{fig:eval_exp}
\end{figure}

Given a task sequence, the network will be initialised and trained with the first task's training data. Then for each subsequent task, the network's output layer will be extended to include its new classes and the network will be trained with the task's training data and in-memory data. The in-memory data are sampled from all the previous tasks' training data. The size of in-memory data is determined by the memory constraint of a particular HAR system, and it will contribute to retaining old knowledge and thus affect the accuracy of old class classification. Figure~\ref{fig:eval_exp} presents an example of task-incremental evaluation on a HAR dataset. As the performance of continual learning can be subject to task sequences, to reduce the bias, we generate several random task sequences (i.e., 30) and report averaged accuracy. 

\subsection{Evaluation Metrics}\label{subsec:metrics}
When training a new task, we compute three types of accuracy. (1) \textit{Base} accuracy -- the accuracy of recognising the activity classes in the first task; (2) \textit{Old} accuracy -- the accuracy of recognising all the old activity classes that have learnt before the current task. Both base and old accuracy will indicate the stability of the model; (3) \textit{New} accuracy -- the accuracy of recognising the new activity classes in the current task, which will indicate the plasticity of the model; and (4) \textit{All} accuracy -- the accuracy of recognising all the activity classes that have learned so far, which will indicate the overall performance of the model. The accuracy is measured in \textit{F1-scores}, which balances precision and recall. As most of the HAR datasets have an imbalanced class distribution, we use \textit{F1-macro} and \textit{F1-micro}~\cite{10.5555/3298483.3298514}. 

% \todo[inline]{Martin: This needs rewording. We have to make clear that this is not the original forgetting measure but draws inspiration from it and has developed in a way that makes it more what we think more understanable. Also Formulas have to add the part that only tasks that have learned previousy are considered}

% \paragraph{Proposed evaluation metric:}
% \todo[]{Martin: we need to check this section against the code formula again. I am pretty sure that its wrong here but I forgot what I said exactly for the formula.}
To understand the retention of knowledge for a given task $j$ at an incremental task $k$, we employ a commonly used measure -- forgetting score (FS)~\cite{Chaudhry2018}.  After a model is trained incrementally till task $k > j$, the forgetting score $f_k^j$ is computed as the difference between the maximum accuracy gained for the task $j$ throughout the learning process. However, this score does not concern with the difficulty level of each task. For example, the accuracy decreasing from 60\% to 40\% on a difficult task suggests more forgetting than the accuracy decreasing from 90\% to 70\% on an easy task. To address this limitation, we propose an adapted forgetting score by normalising it with the best accuracy that can be achieved on a task. That is, 
\begin{align*}
    f_k^j = 1 - \frac{a_{k,j}}{\underset{l \in {1, .., k-1}}{\max} a_{l,j}},
\end{align*}
where $a_{k,j}$ is the accuracy for a previous task $j$ at the current task $k$\footnote{We ignore classes whose maximum accuracy is not greater than 0, avoiding the arithmetic error.}. Finally, the average forgetting $FS_k$ at $k-$th task is normalised against the number of tasks seen previously, i.e.,
\begin{align*}
FS_k=\frac{1}{k-1}\sum_{j=1}^{k-1}f_k^j.    
\end{align*}
 A large $FS$ score implies server forgetting. At the extreme, when $FS$ is 1, it suggests that the knowledge on the old task is completely forgotten.

\subsection{Baseline}\label{subsec:baseline}
Besides the selected models, we consider two baselines that serve as an upper and lower bound of the performance of continual learning. (1) \textbf{Offline batch learning}: we train a network with the training data on all the classes simultaneously, and (2) \textbf{finetuning}: we do not use any holdout samples for replay and only use training data on each new class to update the model with the plain cross-entropy loss. 

\subsection{Architecture Configuration}\label{subsec:architecture}
While the original implementations of our selected techniques use a variety of architectures\footnote{Most of the vision-based methods use standard CNN architectures.} and classifiers\footnote{For instance, iCaRL uses nearest-of-mean classifier.}, we try to maintain a fair comparison premise by using the fully connected networks across all our experiments. To retain the best characteristics of these techniques, we follow the state-of-the-art methodology~\cite{Kemker2018b,Chaudhry2018} to conduct grid search for the network architecture on each dataset. Table \ref{tab:hyperparameters} provides a list of optimised configurations. The initial learning rate (LR) is reduced by a factor of 10 after every scheduler step following the first scheduling epoch. For the choice of the number of neurons in the hidden layers, we rely on fractions of feature dimensions of each dataset. For the particular case of training GEM, we tune the initial learning rate to 0.01 for TWOR and ARUBA, and to 0.001 for DSADS, as a larger learning rate makes it difficult to converge and leads to low accuracy. The scheduling epochs for TWOR and ARUBA are set to 40 and that for DSADS to 50 while the wight decay rates are set to $1e^{-6}, 1e^{-4}$ and $2e^{-6}$ for TWOR, DSADS and ARUBA respectively. 

\begin{table}[!h]
    \centering
    \caption{Hyperparameter configuration for training}
    \label{tab:hyperparameters}
    \includegraphics[width=.95\textwidth]{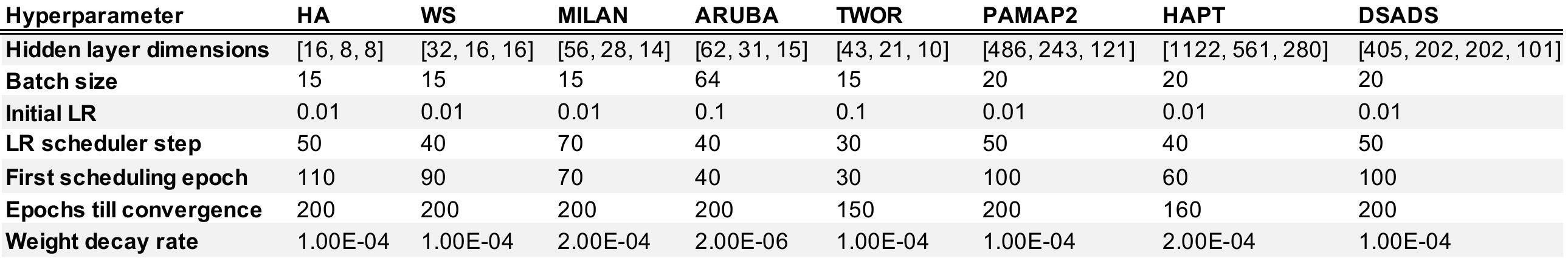}
    % \vspace{-8\baselineskip}
\end{table}

For each comparison technique, we run grid search on their own hyperparameters, and select the best value that leads to the highest accuracy on each dataset. We list the setting in the following. For LwF, the loss balance weight $\lambda_o$ is set as 1.6 for all the datasets. For R-EWC, the regularisation coefficient $\lambda$ in Eq~\ref{eq:ewc} is set as 5 for ARUBA and TWOR and 3 for all the other datasets. For MAS, the regularisation coefficient $\lambda$ in Eq~\ref{eq:mas} is set as 0.1 for TWOR, ARUBA, and MILAN datasets and 0.25 for all the other datasets. 
% \todo[inline]{What is the value for the other datasets}

\section{Results and Analysis}\label{sec:results}
This section presents the results in response to the research questions raised in Section~\ref{sec:experiments}. For each question, we summarise the key results in bold, followed by the analysis. 

\subsection{Overall Comparison}\label{subsec:overall}

\begin{table}[!htbp]
    \centering
    \caption{Comparison of micro-F1 scores on comparison techniques (\footnotesize{The best accuracy is highlighted in bold.})}
    \label{tab:overall_micro}
        \includegraphics[width=\textwidth]{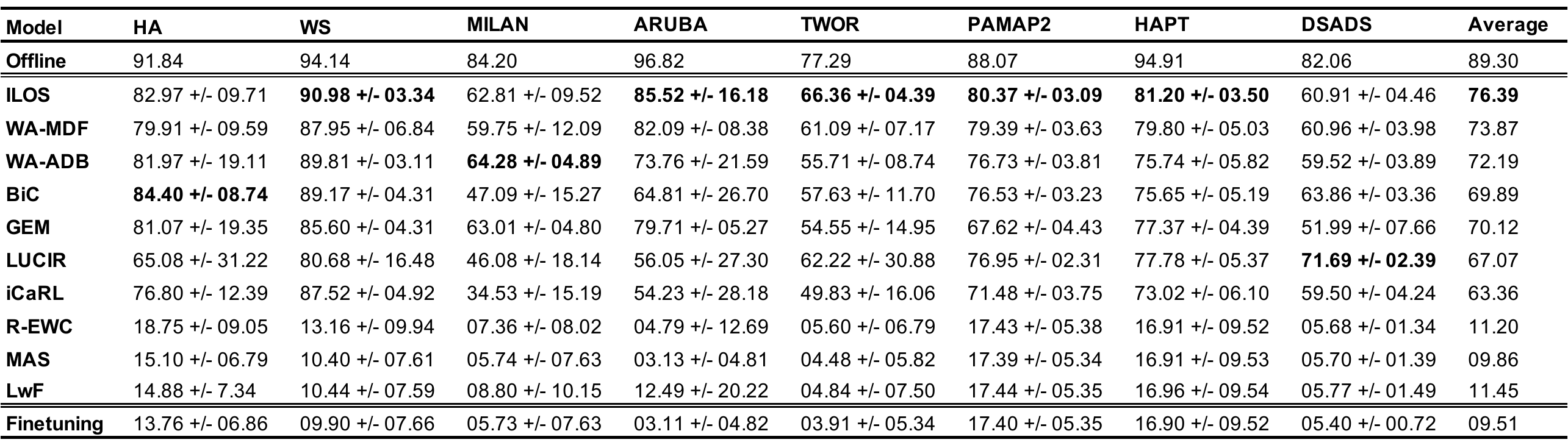}
\end{table}

Table~\ref{tab:overall_micro} and \ref{tab:overall_macro} report the mean and standard deviation of micro- and macro-F1 scores on 10 selected and 2 baseline models across 8 datasets over 30 runs. The last column of these two tables average the accuracy across all the datasets and demonstrates which technique works best.  For all the models that use memory replay, we randomly sample 6 samples from each old class after training each task. 

The offline baseline provides the reference accuracy which implies the difficulty level of each dataset. As shown in Table~\ref{tab:overall_micro} and \ref{tab:overall_macro}, the datasets that have more balanced activity distribution and easier-to-separate classes gain higher micro- and macro-F1 scores; \textit{e.g.}, WS (94.14\% and 87.41\% in micro- and macro-F1) and PAMAP2 (94.91\% and 84.78\% in micro- and macro-F1). The imbalanced datasets often result in higher micro-F1 but lower macro-F1;  \textit{e.g.}, ARUBA (96.82\% and 58.82\% in micro- and macro-F1) and TWOR (77.20\% and 45.35\% in micro- and macro-F1). As lower bound, the finetuning baseline suffers the most catastrophic forgetting, with the overall micro-F1 9.51\% and macro-F1 2.11\% averaged across all the datasets, which are significantly lower than the accuracy achieved from offline. In comparison with the above baselines, we draw the following observations on the selected continual learning techniques.

\begin{table}[!htbp]
    \centering
    \caption{Comparison of macro-F1 scores on comparison techniques (\footnotesize{The best accuracy is highlighted in bold.})}
    \label{tab:overall_macro}
        \includegraphics[width=\textwidth]{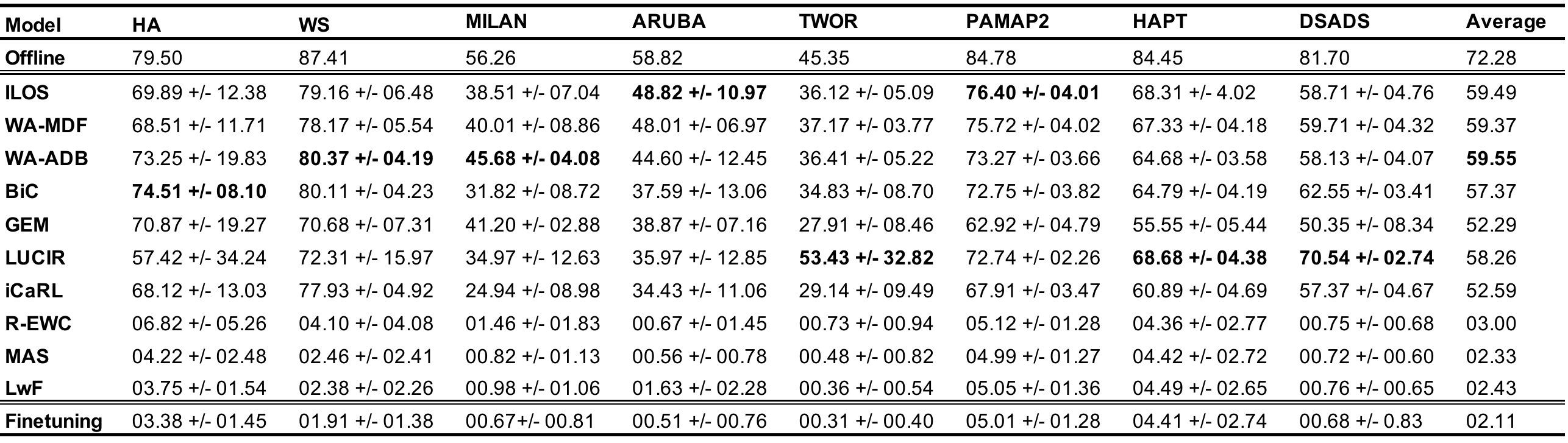}
\end{table}

\noindent\textbf{Rehearsal methods significantly outperform regularisation-alone methods}, as the averaged difference in micro- and macro-F1 is around 50\%. More specifically, from R-EWC to iCaRL, micro-F1 increases from 11.20\% to 63.36\% in Table~\ref{tab:overall_micro}, and macro-F1 from 3\% to 52.59\% in Table~\ref{tab:overall_macro}. Regularisation alone does not help retain the knowledge of the original model as LwF, MAS and EWC only produce roughly the lower bound accuracy. For example, the success of LwF depends on the similarity of new tasks to old tasks. Distribution shift between old and new tasks will result in a discrepancy in the KD loss when predicting the class probability using the old model. The errors are accumulated over incremental learning and will significantly impact its performance~\cite{chen_liu_2018, lange2019continual, He2020IncrementalLI}. In \ac{HAR}, each activity class can have a drastically different sensor feature signature, so in our experiment these regularisation-only methods perform much worse than they do in computer vision experiments. 

\noindent\textbf{Among rehearsal methods,  ILOS, WA-MDF, and WA-ADB perform the best}, improving on the basic rehearsal method iCaRL over 13\% in micro-F1 and 7\% in macro-F1. With memory replay, retraining the model is often affected by the imbalance between a large number of new task's data and a small number of in-memory samples. These methods have effectively avoided optimising the model towards the majority classes and thus better retained the knowledge of the old classes. 

\noindent\textbf{GEM does not perform better than iCaRL}, as it only achieves an increase of 7\% in micro-F1 and the same macro-F1 score as iCaRL. We have also considered the improved version of GEM: A-GEM~\cite{Chaudhry2019}. However, in our experiments, A-GEM often produces worse accuracy than GEM; for example, on PAMAP2 dataset, A-GEM achieves micro-F1 of 37.86\% and macro-F1 of 28.35\%, which is more than 30\% lower than GEM. One possible reason is that activities can have diverse patterns, so when A-GEM down-samples the holdout data, it has an even smaller coverage of space and thus its accuracy is worse than GEM. 
% \todo[inline]{We tried with keeping all of the reserved samples instead of just a subset.. accuracies do not improve - Saurav}

\subsection{Balance between Stability and Plasticity}\label{subsec:stability}
\begin{figure*}[h!]
    \centering
\includegraphics[width=\textwidth]{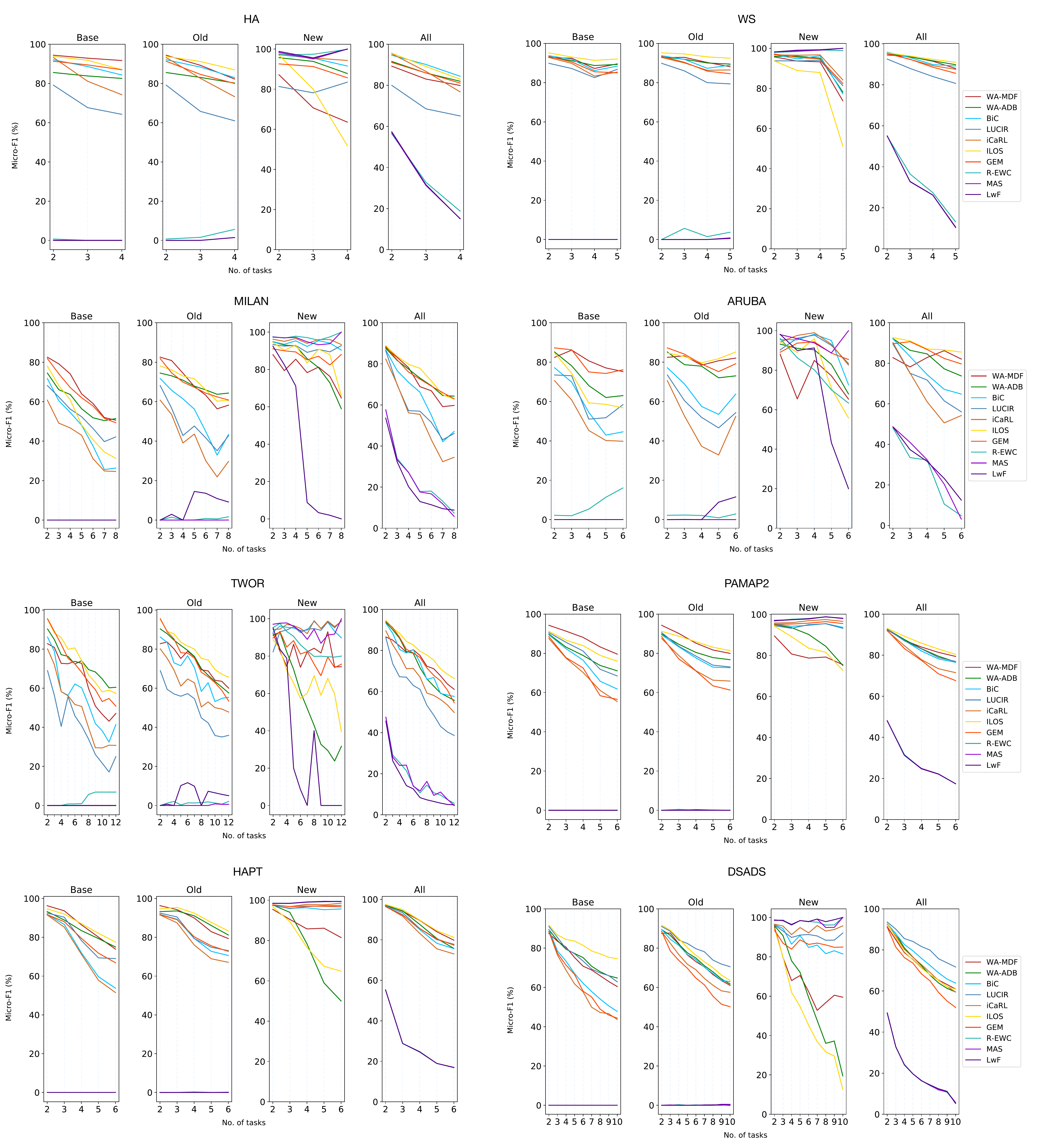}
\caption{Accuracy comparison of task-level performance. It records micro-F1 scores on \textit{Base}, \textit{Old}, \textit{New} and \textit{All} classes.}
    \label{fig:task-level}
\end{figure*}

To look into stability and plasticity over incremental learning, Figure~\ref{fig:task-level} presents task-level micro-F1 of \textit{base}, \textit{old}, \textit{new}, and \textit{all} classes on each technique. As we can see, regularisation methods demonstrate better plasticity as with the increase of tasks, the new accuracy of LwF and EWC remains high. These methods also exhibit poor stability as their base and old accuracy stays at the bottom of the plots, even from the second task on. Overall, the weight alignment methods in the rehearsal category achieve a better balance between stability and plasticity, as the overall accuracy of ILOS, WA-MDF, and WA-ADB suffers less steep drop. Besides, we can observe that each technique has a different performance profile on these 8 datasets, implying the characteristics of the datasets might impact the effectiveness of each technique.

\noindent\textbf{Rehearsal methods that tackle imbalance are less plastic to new classes}. With a direct tuning of the new model's logits based on the previous model, ILOS can help surpass complex regularisation operations in retaining the knowledge but it is not able to quickly adapt to new classes. We also observe that LUCIR performs poorly on new tasks but is robust at preserving old knowledge. An intuitive explanation could be that the design of marginal ranking reinforces the model's confidence at recognising ground truth embedding for old class samples after multiple incremental training steps.

%The accuracy on base and old classes demonstrates the stability of a model; i.e., how well the old knowledge is retained, while the accuracy on new classes represents the plasticity; i.e., how well a model can accumulate new knowledge. 

%\noindent\textbf{Regularisation methods cannot retain knowledge} as without memory replay, R-EWC, MAS, and LwF have very low accuracy on base and old classes, even from the second task on. With respect to rehearsal methods, we can see from Figure~\ref{fig:task-level} that the performance on base and old classes gradually decreases, while GEM and iCaRL have steeper slope. 

\begin{figure}[!htbp]
    \centering
    \includegraphics[width=\textwidth]{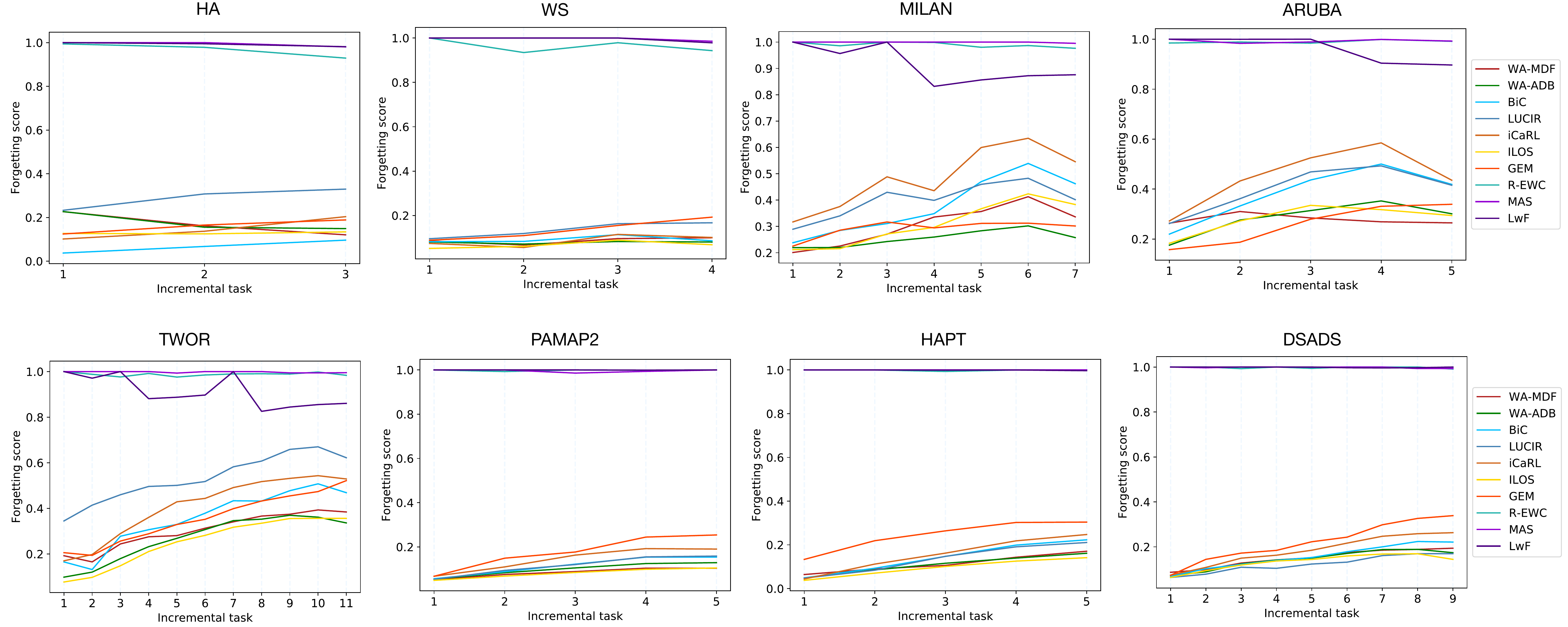}
    \caption{Comparison of forgetting scores on selected techniques}
    \label{fig:forgetting}
\end{figure}

To further inspect the catastrophic forgetting, we present the new forgetting score $FS$ in Figure~\ref{fig:forgetting}. $FS$ on these three regularisation methods is at the upper bound of 1.0 for the first task and stays much higher than the other techniques for the following tasks, suggesting that these techniques suffer an almost immediate total forgetting effect where they cannot come back from. The high forgetting scores of R-EWC conform to the finding of \cite{Kemker2018b} stating that EWC-based methods are poor at learning new categories incrementally.

Among rehearsal-based methods, ILOS and WA-ADB exhibit the best knowledge retention as their averaged $FS$ across the tasks on all the datasets is the lowest; \textit{e.g.}, the old classes only lose 20\% of their best accuracy throughout continuous learning. In contrast, iCaRL and LUCIR are the worst with their $FS$ being 30\%, which is averaged over all the datasets in Figure~\ref{fig:forgetting}. GEM shows a relatively greater tendency of increase in $FS$ as the number of incremental tasks grows. This is evident across the plots for DSADS, ARUBA, TWOR, WS and HA. 

We also observe that after a certain number of incremental tasks, the inertia of forgetting on rehearsal methods is relieved. For instance, iCaRL, BiC, LUCIR and ILOS reach the threshold on ARUBA at the 3rd task while GEM and WA-ADB attain this at the 4th task following which their respective $FS$ witnesses either a plateau or start decreasing. There is no clear correlation between the forgetting effect and the number of tasks being learnt, as it can be dataset-specific, especially the interference between the old and new activities.

\noindent\textbf{There exists a strong effect of the amount and distribution of training data attributes on forgetting}.
For the rehearsal methods, we can see that the datasets that have a long tail distribution ({i.e.}, many activities have low frequency) have much higher forgetting scores. For example in Figure~\ref{fig:data-hist}, TWOR has 19 out of 23 activities whose occurrence is less than 6\%, MILAN has 11 out of 15 activities whose occurrence is less than 8\%, and ARUBA has 8 out of 11 activities whose occurrence is less than 3\%. As shown in Figure~\ref{fig:forgetting}, $F
S$  of the rehearsal based techniques on these datasets is around 15\% higher than those on the other datasets. 

\subsection{Effect of Regularisation}\label{subsec:regularsion_comparison}
From the above results, we can see that regularisation alone does not demonstrate any advantage from a naive finetuning approach, but will their performance be improved when combined with memory replay? If so, which regularisation is more effective in HAR? To answer these two questions, we design the following experiment that uses holdout samples with each regularisation term. More specifically, we look into the following settings: (1) cross-entropy (CE) only (as a baseline without any regularisation) in Eq~4, (2) KD in Eq~\ref{eq:kd_loss}, (3) EWC in Eq~\ref{eq:ewc}, (4) MAS in Eq~\ref{eq:mas}, (5) LUCIR discrimination loss (LUCIR-DIS) $\mathcal{L}^G_{dis}$ in Eq~\ref{eq:lessforget}, (6) LUCIR marginal ranking loss (LUCIR-MR) $\mathcal{L}_{mr}$ in Eq~\ref{eq:lucir-mr}, (7) LUCIR combination loss (LUCIR) in Eq~\ref{eq:LUCIRcomb}, (8) ILOS cross-entropy loss (ILOS-CE) in Eq~\ref{eq:ilos-ce} and (9) ILOS combination loss (ILOS) in Eq~\ref{eq:ilos}. To make a fair comparison, we set up the same setting for each technique, including randomly sampling holdout data from old classes' training data and employing the same training procedure. 

\begin{figure}[!htbp]
    \centering
    \begin{subfigure}[b]{0.95\textwidth}
    \includegraphics[width=0.95\textwidth]{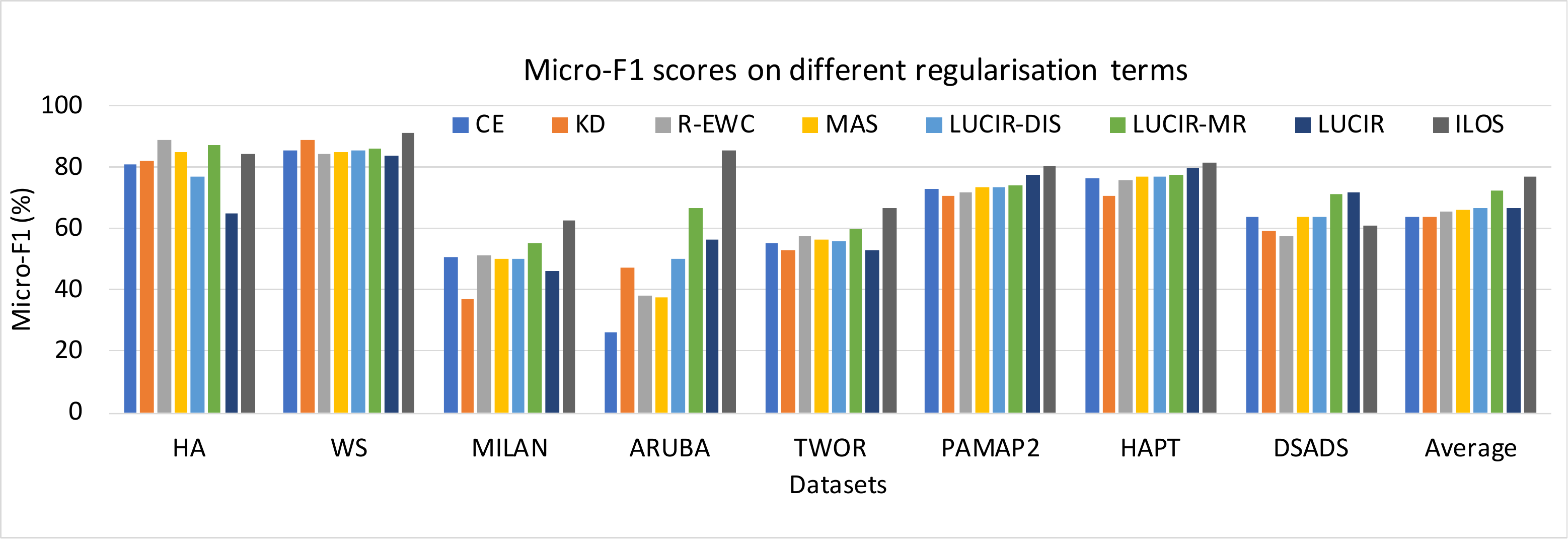}
    % \caption{Comparison of micro-F1 scores on different regularisation terms.}
    % \label{fig:loss_micro}
    \end{subfigure}
    
    \begin{subfigure}[b]{0.95\textwidth}
    \includegraphics[width=0.95\textwidth]{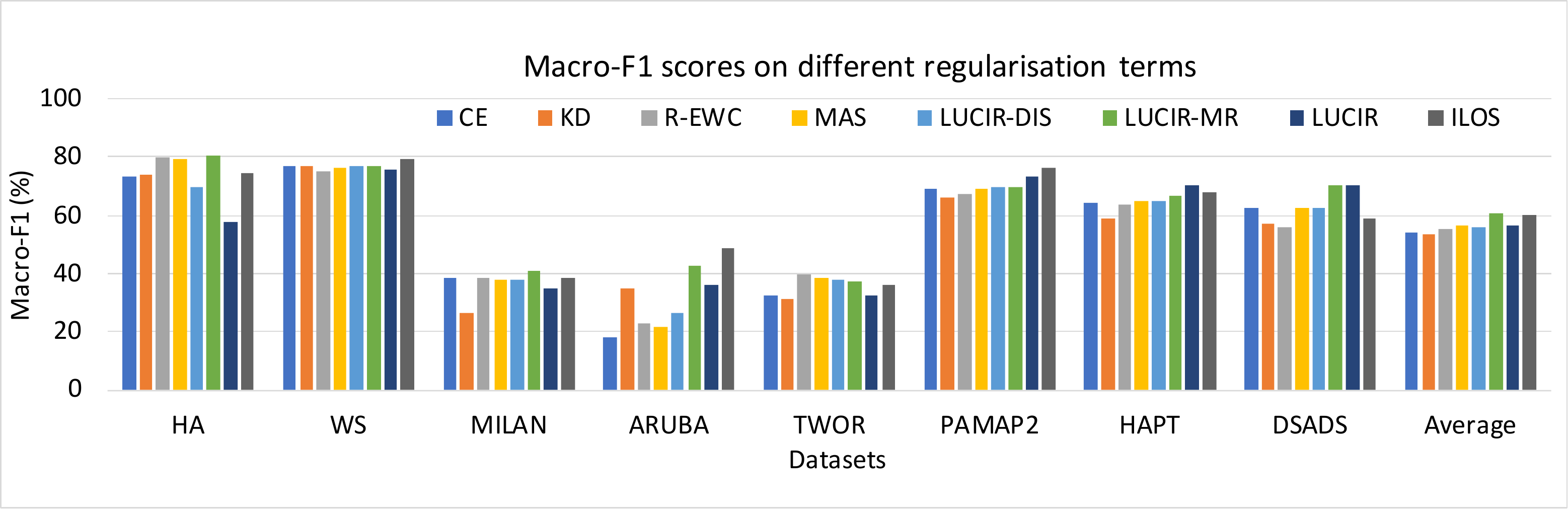}
    % \caption{Comparison of macro-F1 scores on different regularisation terms.}
    % \label{fig:loss_micro}
    \end{subfigure}
    \caption{Performance comparison of different regularisation terms with memory replay}
    \label{fig:loss}
\end{figure}

Figure~\ref{fig:loss} compares micro- and macro-F1 scores of different regularisation terms with memory replay. We draw the following observations. Firstly, LUCIR-MR and ILOS produce better accuracy than a simple cross-entropy loss. As shown in Figure~\ref{fig:loss}, ILOS and LUCIR-MR have produced averaged 77\% and 61\% in micro-F1, which are 13\% and 7\% higher than the plain CE loss. A possible reason is that LUCIR-MR is dedicated to separating classes and ILOS adjusts the logit output to balance the old and new classes. 

Secondly, the other regularisation terms do not improve the results from the CE loss. We can only see 5\% and 3\% improvement in micro- and macro-F1 from the other regularisation terms over CE in Figure~\ref{fig:loss}. It seems that memory replay and the regularisation terms are dealing with the same problem: \textit{interclass discrepancies}. Therefore, there might not be a distinct advantage for regularisation. This finding is consistent with the latest theoretical work~\cite{knoblauch2020optimal}. The terms that decrease the difference between micro- and macro-F1 the most are inhibiting the learning outcome if used in conjunction with memory replay. For example, ILOS and R-EWC have the lowest F1 scores with memory replay. Looking back at Figure~\ref{fig:task-level}, we can see that ILOS is inhibiting new classes from being learned. 
%  \todo[inline]{Martin: add that the difference of macro and micro is more similar to upper boundary}

\subsection{Effect of Sampling}\label{subsec:sampling}
\begin{figure}[!htbp]
    \centering
    \includegraphics[width=0.98\textwidth]{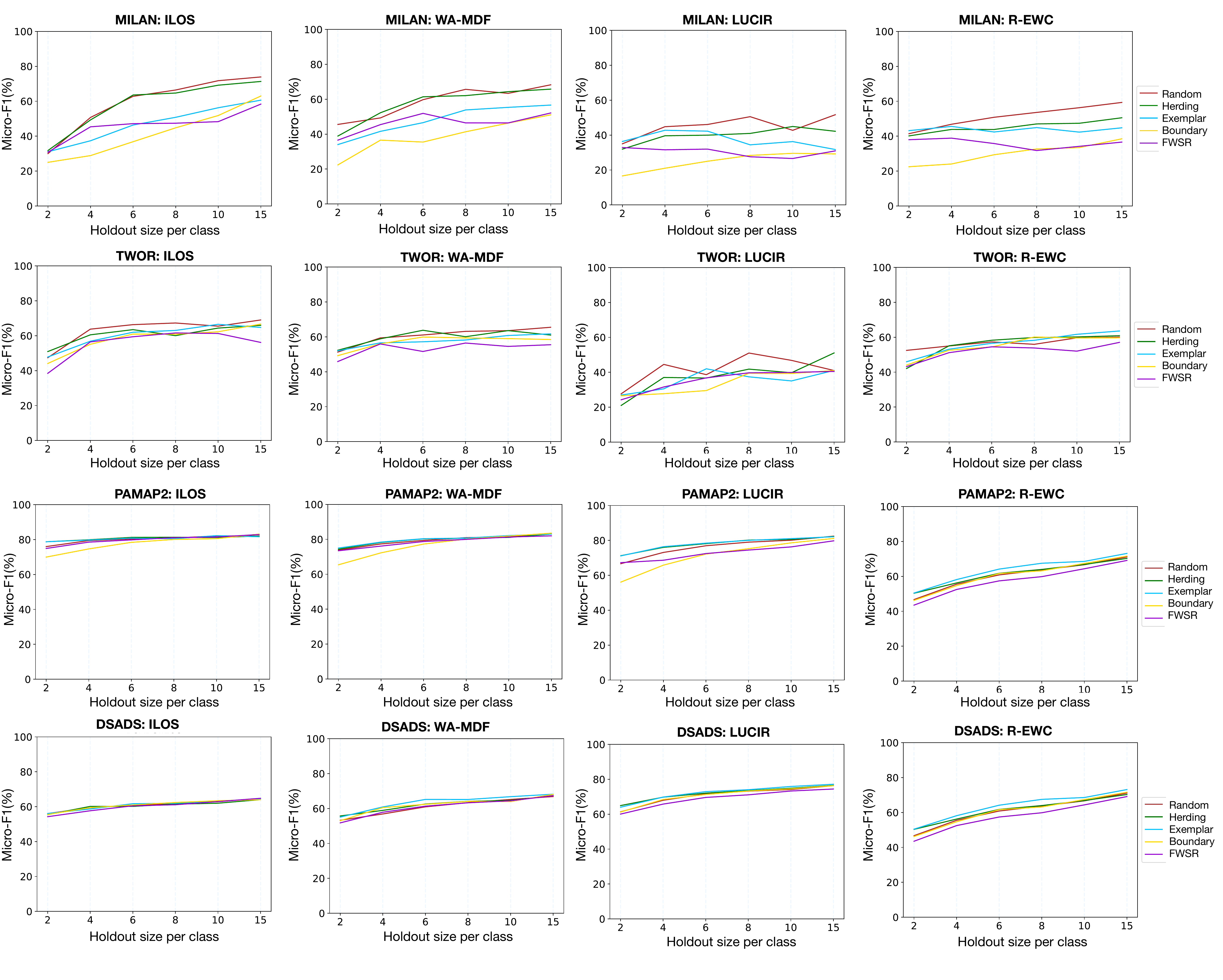}
    \caption{Comparison of accuracy on different sampling techniques and sample sizes. }
    \label{fig:sampling}
\end{figure}

Since rehearsal methods can improve knowledge retention significantly, now the questions are (1) how many samples are needed to store in memory and (2) what sampling strategy is most effective for selecting these samples that are representative for old classes. To investigate these questions, we experiment  different holdout sizes from 2 to 15 with a step size of 2 on widely adopted sampling techniques, including \textit{random} sampling, \textit{herding}~\cite{Welling2009}, \textit{exemplar} sampling, \textit{Frank-Wolfe Sparse Representation} (FWSR) sampling~\cite{Cheng2018}, and \textit{boundary} sampling. Herding is to select the top $s$ samples that are the closest to the mean of each class~\cite{Welling2009}. FWSR sampling selects a subset of the data that effectively describes the entire data set. Exemplar sampling selects the centroid data points of each class. Boundary sampling~\cite{8986833}, a recent sampling technique for incremental learning, selects exemplars on the class decision boundary and overlapping region based on local geometrical and statistical information. Figure~\ref{fig:sampling} compares the micro-F1 scores of sampling techniques on different sample sizes on 4 selected methods and 4 datasets. 

\noindent\textbf{Holdout size does not impact much on the accuracy} in that the accuracy on ILOS, WA-MDF, and LUCIR in Figure~\ref{fig:sampling} does not vary much with the increase of the holdout size. For most of the datasets, the accuracy converges when the holdout size is around 4 or 6. 

\noindent\textbf{In terms of the sampling techniques, random (in red), exemplar (in dark green), and herding (in light green) work better} than the other two more complex techniques, as presented in Figure~\ref{fig:sampling}. Boundary sampling (in orange) works the worst, whose accuracy often stays at the bottom. Sensor data often contains noise and outliers~\cite{CHEN202149}, so it is difficult to characterise geometric shape or precise class boundary of an activity, and thus both boundary and FWSR do not work well. Exemplar and herding try to capture most representative samples, and work well when the holdout size is small; \textit{i.e.}, 2. Random sampling works generally well and is consistent with the results on images~\cite{wen2018fewshot}, as randomness seems to have better coverage in the entirety of data space with minimal bias for certain data distributions.

\begin{figure}[h!]
    \centering
    \includegraphics[width=0.95\textwidth]{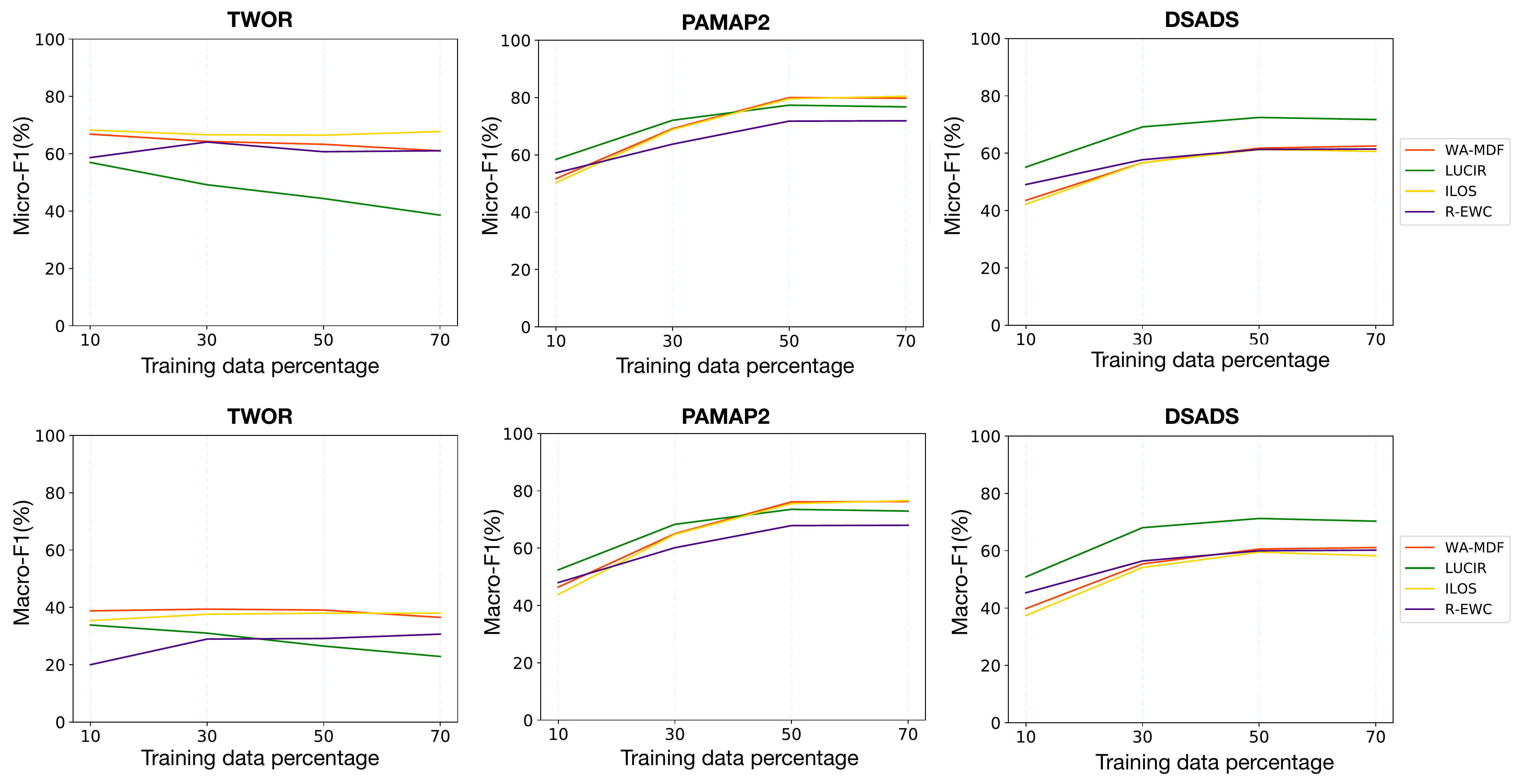}
    \caption{Comparison of accuracy on different training data percentage. \footnotesize{The results demonstrate that the selected lifelong learning techniques are insensitive to training data size.}}
    \label{fig:training}
\end{figure}

\subsection{Effect of Training Data Sizes}\label{subsec:training_data}
Training data for new classes can be more difficult to acquire in a continual learning setting in HAR, as new classes are discovered when the system is already deployed and running. For example, it often relies on users' voluntary self-annotation when the sensor system is deployed in the real world. These experiments are designed to assess the impact of training data size on the selected techniques. We reduce the training data percentage from 70\% to 10\% with a step size of 20\% and report their micro- and macro-F1 in Figure~\ref{fig:training}. The results show that these techniques are not sensitive to training data size, as the accuracy does not change after 30\% training data. For datasets with high imbalance (TWOR), the accuracy of the selected methods does not change, staying at 66\% in micro-F1 and 36\% in macro-F1. The reason is that some of the classes have very little samples; e.g., the activity `R1\_bath' only has 19 samples, and the increase in training percentage still does not lead to many samples. We observe that LUCIR decreases accuracy with the increase in training data. The reason is that TWOR has many difficult-to-separate activities, which makes it more challenging to find good anchors when more training data is available. 

% However, the increasing imbalance due to larger train sizes actually hinders the overall performance over smaller sizes. The margin ranking loss of LUCIR is specially sensitive to this across both datasets.
\subsection{Computation Cost Analysis}\label{subsec:cost}
Computation cost is an important consideration in HAR since devices are usually under memory and computational power constraints. Table~\ref{tab:CompCost} presents the computation time of all the selected methods on training a single incremental task and as well as the whole task sequence. For each dataset, we highlight the technique with the longest training time. We report the averaged training time in the last column for the overall comparison. All the training is performed on a modest computer with Intel Core i5 8400, 32GB memory, and 2$\times$500GB SSD.

\begin{table}[!hbtp] % use tables like this
\centering
\caption[-]{Comparison of computation time (in seconds) of training each incremental (\textit{Incre.}) task and all the tasks (\textit{Total}). (\footnotesize{The longest training time is highlighted in bold.})}
\label{tab:CompCost}
\includegraphics[width=0.9\textwidth]{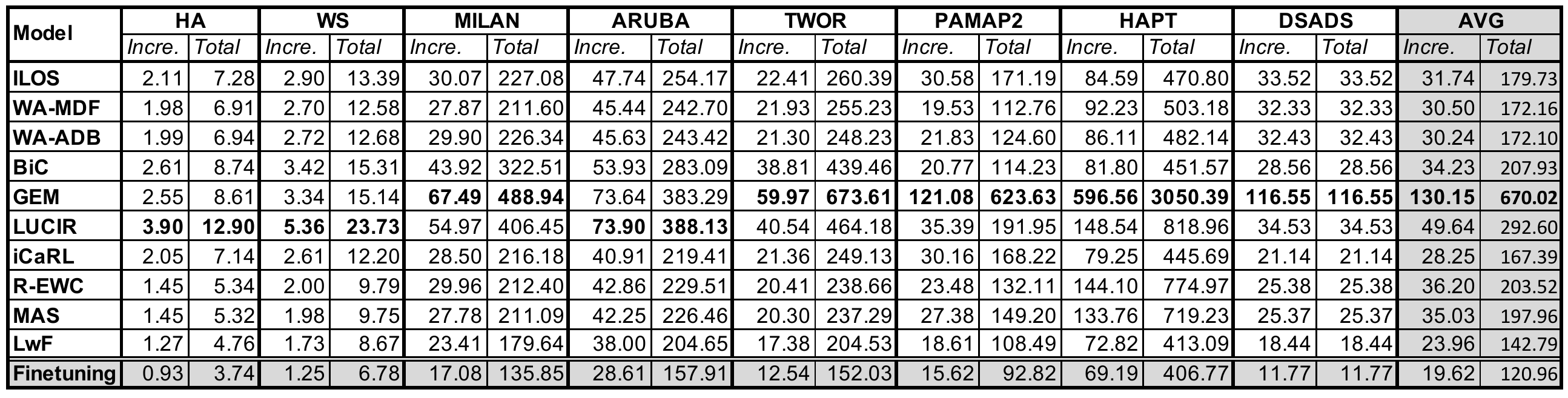}

\end{table}

As shown in Table~\ref{tab:CompCost}, \noindent\textbf{GEM and LUCIR are the most expensive ones} among the others, especially the incremental training time of GEM (i.e., 130s) is 6 times higher than finetuning on average (i.e., 20s). This is because since the quadratic optimisation in  GEM is computationally expensive. LUCIR needs to find good anchors, which incurs an extra cost, so it takes significantly longer training time than the other 8 techniques. The rehearsal based methods take longer to compute than the methods purely based on regularisation. For example, LwF, as the simplest technique, is the least computationally expensive; that is, 20s longer in total training time than finetuning on average. Comparing the weight alignment techniques, BiC is more expensive as it requires to tune the bias correction layer, which happens after a new task has converged. This is slightly more expensive (i.e., 30s more in total training time) than the relatively simple correction methods of WA-MDF/ADB.

\begin{table}[!htbp]
    \centering
    \caption{Summary of selected techniques}
    \label{tab:summary}
    \includegraphics[width=0.9\textwidth]{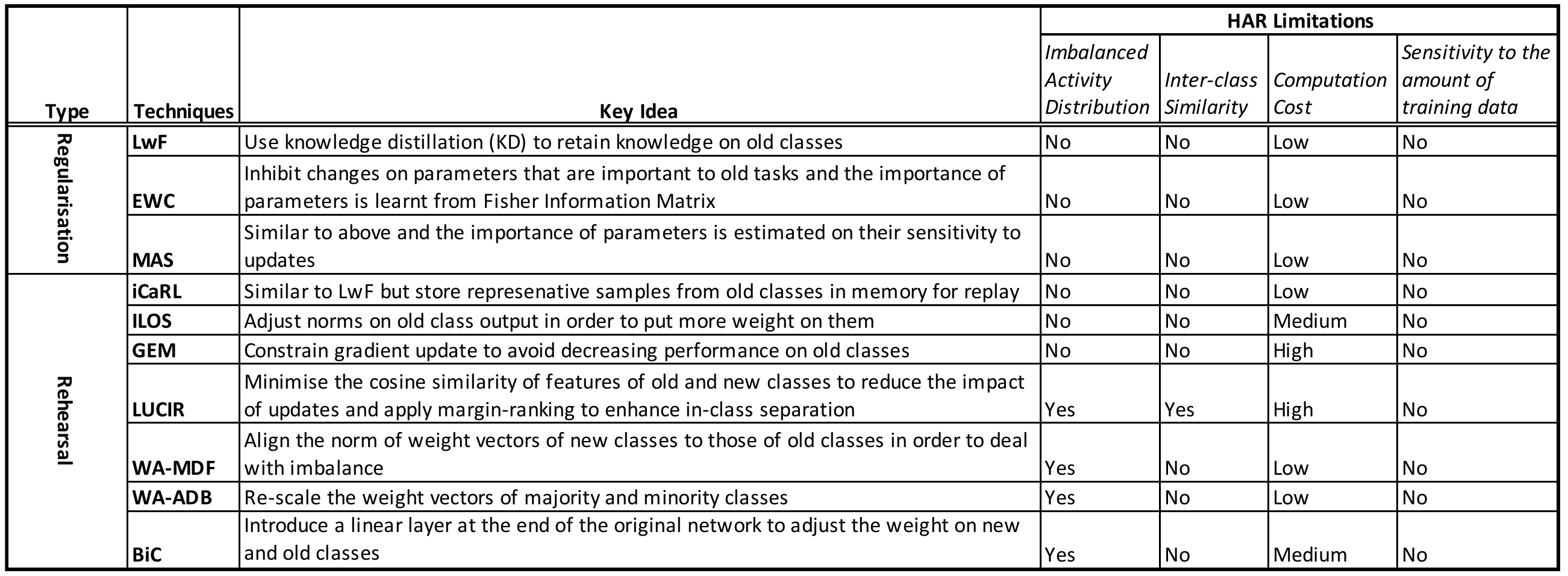}
\end{table}

\section{Discussion}~\label{sec:discussion}
This section summarises the above results and provides guidelines on what type of techniques to use under what conditions for HAR systems. Table~\ref{tab:summary} lists the key ideas of each technique and how well they tackle \ac{HAR} limitations.

\subsection{Imbalanced Activity Distribution}
Among the selected techniques, ILOS and WA-ADB produce higher macro-F1 scores than the others on more skewed datasets such as WS, MILAN and ARUBA. WA-ADB accounts for the ratio of training samples in each class at each incremental training step. For ILOS, the averaging of output logits reduces the magnitude of the logits of new dominant classes.   

\subsection{Inter-class Similarity}
Inter-class similarity is the key characteristic of HAR. Some activities can have very similar sensor signatures. For example in the DSADS dataset, the activity classes between `lying on right' and `lying on back' exhibit high correlation, which can result in overlapping decision boundaries. LUCIR and GEM have attempted to tackle this problem, and LUCIR outperforms GEM with 6\% increase in macro-F1. In addition, LUCIR has achieved the highest accuracy on TWOR dataset that faces the largest challenge of inter-class separation; that is, distinguishing meal preparation from one resident to another. A promising future direction in HAR is to introduce regularisation terms that enhance the discriminability of the network; {e.g.}, the contrastive loss~\cite{Dai2019ContrastivelySC} that is dedicated to learning the difference between two sets of data. 

\subsection{Computation Cost}
Except for GEM and LUCIR, the other selected methods take about twice of training time as finetuning. This is acceptable for most modern devices that run human activity recognition. In terms of memory requirements, as the performance of these methods is not sensitive to the holdout size, so we advise a small number ({i.e.}, 2 or 4 data points per class) will be sufficient. Note that most of the regularisation methods require to store the network parameters in memory to compare before and after an update. For a large neural network that consists of a large number of parameters, this might even be considerably more costly than keeping samples. Also, in our experiments, there is no advantage of sophisticated sampling techniques over random sampling of our datasets. 

% there is no need for complex algorithms for sampling holdout data in  

% Looking back at tables \ref{tab:overall_micro} and \ref{tab:overall_macro} one can see that performance is not necessarily linked to higher computation time as ILOS usually good performance does not take long to compute. It is also worth mentioning, that LUCIR's and GEM's performance is falling behind their computational cost. The same can be said about BiC, since its results are generally on par or worse than the other weight alignment methods even though the computation time is higher.\\
% Generally one can say that even though the computation cost of the memory replay methods is higher than the regularization techniques', this is well effort well spent as the performance improvement makes up for it.\\
% In terms of memory requirements, an conclusion is not very easy to make since it is very much dependant on how many samples one chooses for the holdout and also how big the underlying model is. A larger holdout will naturally lead to more memory consumption for the replay methods but one should not forget that the regularization methods need to keep a copy of the network parameters in memory to compare the parameters before and after an update. Since modern deep learning models can consist of very large amounts of parameters, this might even be considerably more costly than keeping samples.\\
% If one wants the best bang for the buck, ILOS, WA-MDF or WA-ADB might be the first techniques to try.
% \todo[]{MArtin: we maybe want to put the end where recommendation is given in a separate section}

\subsection{Scarcity of Labelled Data}
The selected techniques are not sensitive to training data size, which is good for incremental learning when labelling new activities is even more difficult. In our experiments, 30\% of the training data will be sufficient. However, given a large dataset, 30\% still means a large number of samples; e.g., DSADS needs to label 144 samples per class. In the future, we could look into few-shot learning algorithms to further reduce the number of training samples on new classes. 

\section{Conclusion and Future Work}\label{sec:conclusion}
This paper presents a comprehensive, empirical evaluation of recent continual learning techniques in a task-incremental setting. We seek answers to essential research questions in \ac{HAR}. We find that rehearsal techniques will lead to the best performance on most of the selected \ac{HAR} datasets. They can be computationally cheap and do not require much memory space. Sophisticated regularisation terms or gradient updates fall short on their promises. The regularisation terms that help to deal with imbalances and inter-class separation achieve more promising accuracy on HAR datasets. In the following, we will focus on the future developments that HAR techniques have to have.

The state-of-the-art continual learning techniques have paved a promising future for continual learning in HAR. However, most of these techniques are either set in a non-realistic continual learning scenario such as permutation MNIST~\cite{ven2019scenarios}, or in scenarios with distinct task boundaries. We envision a continual learning system in a real-world HAR deployment where new activities can occur spontaneously and interweave with the old activities. Therefore, it needs to be able to discover new activities first. This requires to combine the existing continual learning techniques with techniques for anomaly detection; {i.e.}, detecting whether the current sensor data conforms to any existing activity pattern. In HAR, new activity discovery has been extensively studied~\cite{8903481,ye2016discovery,Gjoreski:2017:UOA:3123021.3123044}. The question is how to combine these two types of techniques and form a feedback loop, where a new activity is discovered and fed into a network for extension in an automated fashion without any human intervention. 

Furthermore, acquiring annotations on new tasks can be challenging, as they rely on user input, which can be unavailable or imprecise~\cite{10.1145/1149941.1149967}. Unlike the existing scenario for most of the continual learning techniques, where the training data is abundant (e.g., 1000 or 2000 examples per class) and well-annotated. In HAR, research interest lies into how to obtain annotations and how to produce robust HAR system in the face of imprecise, insufficient annotation~\cite{8796471}. 

Due to sensor degradation, sensor readings will drift over time, and users' behaviour patterns may change due to their health condition. Both will lead to changes in the distributions of the activities over time. This can be considered as \textit{concept drift}, a common problem in streaming data. This adds complication when there is a need for not only learning new tasks but also adapting the model on old tasks. A recent approach that provides a self-constructing methodology to extract hidden layers and neurons from streaming data demonstrates its effectiveness in learning non-stationary data~\cite{PRATAMA2019150}.

In the future, we believe that given the promising results on the existing continual learning techniques, the key challenge for \ac{HAR} can move on to a system-level approach: how to discover new tasks where there is no clear task boundary, deal with sensor noise and concept drift, and more importantly, tackle noisy and scarce annotations on datasets. 

%Lastly, we identify further directions for continual learning in \ac{HAR}.
% \todo[]{MArtin: This sentence is a little clunky. What is it supposed to mean exactly? That the limited memory in har is still enough to do memory replay?} 
% We found that the impact of in-memory sample size on the performance of these techniques is not significant, so the devices where a \ac{HAR} system is deployed can afford the memory space; \textit{i.e.}, a couple of samples per class and each sample has a dimension not greater than 500. We also observed that most of these techniques are not sensitive to the size of training data, which is promising since well-annotated training data is difficult to acquire on \ac{HAR} systems. 
% \todo[inline]{Martin: It feels like there is stuff missing from our results. We should come back to this when all key results are taken into account in the main part and tick them off here.}

% The state-of-the-art techniques have made significant progress in the computer vision domain, however, there exists a room to improve in \todo[]{Martin: Do we wanna talk about the results for \ac{HAR} being worse than vision so limited generalisation capabilities?} enabling continual learning in \ac{HAR}.

% \subsection{Task-free Scenario}

% \subsection{Annotation Scarcity}

% \subsection{Concept Drift}
 
\nocite{10.1016/j.asoc.2015.10.027}
\bibliography{2020-IS-benchmark.bib}

\begin{thebibliography}{10}

\bibitem{ABRAHAM200573}
Wickliffe~C. Abraham and Anthony Robins.
\newblock Memory retention -- the synaptic stability versus plasticity dilemma.
\newblock {\em Trends in Neurosciences}, 28(2):73 -- 78, 2005.

\bibitem{AFONSO2019282}
Luis~C.S. Afonso, Gustavo~H. Rosa, Clayton~R. Pereira, Silke~A.T. Weber,
  Christian Hook, Victor Hugo~C. Albuquerque, and Joao~P. Papa.
\newblock A recurrence plot-based approach for parkinson's disease
  identification.
\newblock {\em Future Generation Computer Systems}, 94:282 -- 292, 2019.

\bibitem{Ahmed5255}
Mobyen~Uddin Ahmed, Staffan Brickman, Alexander Dengg, Niklas Fasth, Marko
  Mihajlovic, and Jacob Norman.
\newblock A machine learning approach to classify pedestrians’ event based on
  imu and gps.
\newblock {\em International Journal of Artificial Intelligence}, 16(2), May
  2020.

\bibitem{Aljundi2018}
Rahaf Aljundi, Francesca Babiloni, Mohamed Elhoseiny, Marcus Rohrbach, and
  Tinne Tuytelaars.
\newblock {Memory Aware Synapses: Learning What (not) to Forget}.
\newblock {\em ECCV}, 2018.

\bibitem{expertgate}
Rahaf Aljundi, Punarjay Chakravarty, and Tinne Tuytelaars.
\newblock Expert gate: Lifelong learning with a network of experts.
\newblock In {\em CVPR 2017}, pages 7120--7129, 07 2017.

\bibitem{Altun:2010:HAR:1881331.1881338}
Kerem Altun and Billur Barshan.
\newblock Human activity recognition using inertial/magnetic sensor units.
\newblock In {\em HBU'10}, pages 38--51, Berlin, Heidelberg, 2010.
  Springer-Verlag.

\bibitem{ALTUN20103605}
Kerem Altun, Billur Barshan, and Orkun Tunael.
\newblock Comparative study on classifying human activities with miniature
  inertial and magnetic sensors.
\newblock {\em Pattern Recognition}, 43(10):3605 -- 3620, 2010.

\bibitem{ALVAREZALVAREZ2013162}
Alberto Alvarez-Alvarez, Jose~M. Alonso, and Gracian Trivino.
\newblock Human activity recognition in indoor environments by means of fusing
  information extracted from intensity of wifi signal and accelerations.
\newblock {\em Information Sciences}, 233:162 -- 182, 2013.

\bibitem{10.1007/978-3-030-01258-8_15}
Francisco~M. Castro, Manuel~J. Mar{\'i}n-Jim{\'e}nez, Nicol{\'a}s Guil,
  Cordelia Schmid, and Karteek Alahari.
\newblock End-to-end incremental learning.
\newblock In Vittorio Ferrari, Martial Hebert, Cristian Sminchisescu, and Yair
  Weiss, editors, {\em Computer Vision -- ECCV 2018}, pages 241--257, Cham,
  2018. Springer International Publishing.

\bibitem{Chaudhry2018}
Arslan Chaudhry, Puneet~K. Dokania, Thalaiyasingam Ajanthan, and Philip~H.S.
  Torr.
\newblock {Riemannian Walk for Incremental Learning: Understanding Forgetting
  and Intransigence}.
\newblock In {\em Lecture Notes in Computer Science (including subseries
  Lecture Notes in Artificial Intelligence and Lecture Notes in
  Bioinformatics)}, 2018.

\bibitem{Chaudhry2019}
Arslan Chaudhry, Ranzato Marc'Aurelio, Marcus Rohrbach, and Mohamed Elhoseiny.
\newblock {Efficient lifelong learning with A-GEM}.
\newblock In {\em 7th International Conference on Learning Representations,
  ICLR 2019}, 2019.

\bibitem{Chen2020DeepLF}
Kaixuan Chen, Dalin Zhang, L.~Yao, Bin Guo, Z.~Yu, and Y.~Liu.
\newblock Deep learning for sensor-based human activity recognition: Overview,
  challenges and opportunities.
\newblock {\em ArXiv}, abs/2001.07416, 2020.

\bibitem{CHEN202149}
Lingqiang Chen, Guanghui Li, and Guangyan Huang.
\newblock A hypergrid based adaptive learning method for detecting data faults
  in wireless sensor networks.
\newblock {\em Information Sciences}, 553:49 -- 65, 2021.

\bibitem{net2net}
Tianqi Chen, Ian Goodfellow, and Jonathon Shlens.
\newblock Net2net: Accelerating learning via knowledge transfer.
\newblock {\em arXiv e-prints}, page arXiv:1511.05641, 11 2015.

\bibitem{chen_liu_2018}
Zhiyuan Chen and Bing Liu.
\newblock {\em Lifelong Machine Learning}.
\newblock Morgan et Claypool Publishers, 2018.

\bibitem{Cheng2018}
Gary Cheng, Armin Askari, Kannan Ramchandran, and Laurent~El Ghaoui.
\newblock {Greedy Frank-Wolfe Algorithm for Exemplar Selection}.
\newblock {\em arXiv e-prints}, nov 2018.

\bibitem{cook09dataset}
D.~Cook and M.~Schmitter-Edgecombe.
\newblock Assessing the quality of activities in a smart environment.
\newblock {\em Methods of Information in Medicine}, 48:480--485, 2009.

\bibitem{Courbariaux2015BinaryConnectTD}
Matthieu Courbariaux, Yoshua Bengio, and Jean-Pierre David.
\newblock Binaryconnect: Training deep neural networks with binary weights
  during propagations.
\newblock {\em ArXiv}, abs/1511.00363, 2015.

\bibitem{Dai2019ContrastivelySC}
Shuyang Dai, Yu~Cheng, Yizhe Zhang, Zhe Gan, Jingjing Liu, and Lawrence Carin.
\newblock Contrastively smoothed class alignment for unsupervised domain
  adaptation.
\newblock {\em ArXiv}, abs/1909.05288, 2019.

\bibitem{8986833}
S.~{Dang}, Z.~{Cao}, Z.~{Cui}, Y.~{Pi}, and N.~{Liu}.
\newblock Class boundary exemplar selection based incremental learning for
  automatic target recognition.
\newblock {\em IEEE Transactions on Geoscience and Remote Sensing}, pages
  1--11, 2020.

\bibitem{dhar2018learning}
Prithviraj Dhar, Rajat~Vikram Singh, Kuan-Chuan Peng, Ziyan Wu, and Rama
  Chellappa.
\newblock Learning without memorizing.
\newblock In {\em CVPR 2019}, 2019.

\bibitem{8667728}
L.~{Fang}, J.~{Ye}, and S.~{Dobson}.
\newblock Discovery and recognition of emerging human activities using a
  hierarchical mixture of directional statistical models.
\newblock {\em IEEE Transactions on Knowledge and Data Engineering}, 2019.

\bibitem{farajtabar2019orthogonal}
Mehrdad Farajtabar, Navid Azizan, Alex Mott, and Ang Li.
\newblock Orthogonal gradient descent for continual learning.
\newblock In {\em AISTATS 2020}, Palermo, Italy, 2020.

\bibitem{farquhar2018robust}
Sebastian Farquhar and Yarin Gal.
\newblock Towards robust evaluations of continual learning.
\newblock In {\em LLARLA 2018}, 2018.

\bibitem{10.5555/3298483.3298514}
Kyle~D. Feuz and Diane~J. Cook.
\newblock Modeling skewed class distributions by reshaping the concept space.
\newblock In {\em AAAI-17}, page 1891–1897. AAAI Press, 2017.

\bibitem{10.1016/j.asoc.2015.10.027}
Fernando Gaxiola, Patricia Melin, Fevrier Valdez, Juan~R. Castro, and Oscar
  Castillo.
\newblock Optimization of type-2 fuzzy weights in backpropagation learning for
  neural networks using gas and pso.
\newblock {\em Appl. Soft Comput.}, 38(C):860–871, January 2016.

\bibitem{Gjoreski:2017:UOA:3123021.3123044}
Hristijan Gjoreski and Daniel Roggen.
\newblock Unsupervised online activity discovery using temporal behaviour
  assumption.
\newblock In {\em ISWC 2017}, pages 42--49, 2017.

\bibitem{10.1145/1149941.1149967}
Frank~Allan Hansen.
\newblock Ubiquitous annotation systems: Technologies and challenges.
\newblock In {\em Proceedings of the Seventeenth Conference on Hypertext and
  Hypermedia}, HYPERTEXT '06, page 121–132, New York, NY, USA, 2006.
  Association for Computing Machinery.

\bibitem{REMIND}
Tyler~L. Hayes, Kushal Kafle, Robik Shrestha, Manoj Acharya, and Christopher
  Kanan.
\newblock Remind your neural network to prevent catastrophic forgetting.
\newblock In {\em ECCV 2020}, 2020.

\bibitem{He2020IncrementalLI}
Jiangpeng He, Runyu Mao, Zeman Shao, and Fengqing Zhu.
\newblock Incremental learning in online scenario.
\newblock {\em ArXiv}, abs/2003.13191, 2020.

\bibitem{Hinton2015}
Geoffrey Hinton, Oriol Vinyals, and Jeff Dean.
\newblock {Distilling the Knowledge in a Neural Network}.
\newblock {\em arXiv e-prints}, mar 2015.

\bibitem{Hou2019}
Saihui Hou, Xinyu Pan, Chen~Change Loy, Zilei Wang, and Dahua Lin.
\newblock {Learning a unified classifier incrementally via rebalancing}.
\newblock In {\em CVPR 2019}, 2019.

\bibitem{hung2019compacting}
Ching-Yi Hung, Cheng-Hao Tu, Cheng-En Wu, Chien-Hung Chen, Yi-Ming Chan, and
  Chu-Song Chen.
\newblock Compacting, picking and growing for unforgetting continual learning.
\newblock In {\em NeurIPS}, pages 13647--13657, 2019.

\bibitem{Huszr2018NoteOT}
Ferenc Husz{\'a}r.
\newblock Note on the quadratic penalties in elastic weight consolidation.
\newblock {\em Proceedings of the National Academy of Sciences of the United
  States of America}, 115 11:E2496--E2497, 2018.

\bibitem{IRFAN202180}
Muhammad Irfan, Zheng Jiangbin, Muhammad Iqbal, and Muhammad~Hassan Arif.
\newblock A novel lifelong learning model based on cross domain knowledge
  extraction and transfer to classify underwater images.
\newblock {\em Information Sciences}, 552:80 -- 101, 2021.

\bibitem{Jung2020AGS}
Sangwon Jung, Hongjoon Ahn, Sungmin Cha, and Taesup Moon.
\newblock Continual learning with node-importance based adaptive group sparse
  regularization.
\newblock In {\em NeurIPS 2020}, 2020.

\bibitem{Kemker2018b}
Ronald Kemker, Marc McClure, Angelina Abitino, Tyler~L. Hayes, and Christopher
  Kanan.
\newblock Measuring catastrophic forgetting in neural networks.
\newblock In {\em AAAI-18}, pages 3390--3398. {AAAI} Press, 2018.

\bibitem{Kim2019AdjustingDB}
B.~{Kim} and J.~{Kim}.
\newblock Adjusting decision boundary for class imbalanced learning.
\newblock {\em IEEE Access}, 8:81674--81685, 2020.

\bibitem{Kirkpatrick2016}
James Kirkpatrick, Razvan Pascanu, Neil Rabinowitz, Joel Veness, Guillaume
  Desjardins, Andrei~A. Rusu, Kieran Milan, John Quan, Tiago Ramalho, Agnieszka
  Grabska-Barwinska, Demis Hassabis, Claudia Clopath, Dharshan Kumaran, and
  Raia Hadsell.
\newblock {Overcoming catastrophic forgetting in neural networks}.
\newblock {\em arXiv e-prints}, dec 2016.

\bibitem{knoblauch2020optimal}
Jeremias Knoblauch, Hisham Husain, and Tom Diethe.
\newblock Optimal continual learning has perfect memory and is {NP}-hard.
\newblock In Hal~Daumé III and Aarti Singh, editors, {\em Proceedings of the
  37th International Conference on Machine Learning}, volume 119, pages
  5327--5337, Virtual, 13--18 Jul 2020. PMLR.

\bibitem{Krizhevsky09learningmultiple}
Alex Krizhevsky.
\newblock Learning multiple layers of features from tiny images.
\newblock {\em Handbook of Systemic Autoimmune Diseases}, 1(4), 2009.

\bibitem{lange2019continual}
Matthias~De Lange, Rahaf Aljundi, Marc Masana, Sarah Parisot, Xu~Jia, Ales
  Leonardis, Gregory Slabaugh, and Tinne Tuytelaars.
\newblock A continual learning survey: Defying forgetting in classification
  tasks.
\newblock {\em arXiv preprint arXiv:1909.08383}, 2019.

\bibitem{lesort2019continual}
Timoth{\'e}e Lesort, V.~Lomonaco, A.~Stoian, D.~Maltoni, David Filliat, and
  N.~Rodr{\'i}guez.
\newblock Continual learning for robotics: Definition, framework, learning
  strategies, opportunities and challenges.
\newblock {\em Inf. Fusion}, 58:52--68, 2020.

\bibitem{Li2018}
Z.~{Li} and D.~{Hoiem}.
\newblock Learning without forgetting.
\newblock {\em IEEE Transactions on Pattern Analysis and Machine Intelligence},
  40(12):2935--2947, 2018.

\bibitem{Li2016}
Zhizhong Li and Derek Hoiem.
\newblock {Learning Without Forgetting}.
\newblock In {\em ECVC 2016}, pages 614--629, 2016.

\bibitem{LIU201641}
Li~Liu, Yuxin Peng, Shu Wang, Ming Liu, and Zigang Huang.
\newblock Complex activity recognition using time series pattern dictionary
  learned from ubiquitous sensors.
\newblock {\em Information Sciences}, 340-341:41 -- 57, 2016.

\bibitem{LIU201713}
Li~Liu, Shu Wang, Guoxin Su, Bin Hu, Yuxin Peng, Qingyu Xiong, and Junhao Wen.
\newblock A framework of mining semantic-based probabilistic event relations
  for complex activity recognition.
\newblock {\em Information Sciences}, 418-419:13 -- 33, 2017.

\bibitem{xialei2018forgetting}
X.~{Liu}, M.~{Masana}, L.~{Herranz}, J.~{Van de Weijer}, A.~M. {López}, and
  A.~D. {Bagdanov}.
\newblock Rotate your networks: Better weight consolidation and less
  catastrophic forgetting.
\newblock In {\em 2018 24th International Conference on Pattern Recognition
  (ICPR)}, pages 2262--2268, 2018.

\bibitem{Lopez-Paz2017b}
David Lopez-Paz and Marc'Aurelio Ranzato.
\newblock Gradient episodic memory for continual learning.
\newblock In {\em Proceedings of the 31st International Conference on Neural
  Information Processing Systems}, page 6470–6479, Red Hook, NY, USA, 2017.
  Curran Associates Inc.

\bibitem{LYU2021454}
Gengyu Lyu, Songhe Feng, and Yidong Li.
\newblock Noisy label tolerance: A new perspective of partial multi-label
  learning.
\newblock {\em Information Sciences}, 543:454 -- 466, 2021.

\bibitem{MacKay1992}
David J.~C. MacKay.
\newblock A practical bayesian framework for backpropagation networks.
\newblock {\em Neural Comput.}, 4(3):448–472, May 1992.

\bibitem{10.3389/frai.2020.00019}
Jaya~Krishna Mandivarapu, Blake Camp, and Rolando Estrada.
\newblock Self-net: Lifelong learning via continual self-modeling.
\newblock {\em Frontiers in Artificial Intelligence}, 3:19, 2020.

\bibitem{Merolla2016DeepNN}
Paul Merolla, Rathinakumar Appuswamy, John~V. Arthur, Steven~K. Esser, and
  Dharmendra~S. Modha.
\newblock Deep neural networks are robust to weight binarization and other
  non-linear distortions.
\newblock {\em ArXiv}, abs/1606.01981, 2016.

\bibitem{MROZEK2020132}
Dariusz Mrozek, Anna Koczur, and Bożena Małysiak-Mrozek.
\newblock Fall detection in older adults with mobile iot devices and machine
  learning in the cloud and on the edge.
\newblock {\em Information Sciences}, 537:132 -- 147, 2020.

\bibitem{Ostapenko_Puscas2019}
Oleksiy Ostapenko, Mihai Puscas, Tassilo Klein, Patrick J\"{a}hnichen, and Moin
  Nabi.
\newblock Learning to remember: A synaptic plasticity driven framework for
  continual learning.
\newblock In {\em CVPR '19}, 2019.

\bibitem{PARISI201954}
German~I. Parisi, Ronald Kemker, Jose~L. Part, Christopher Kanan, and Stefan
  Wermter.
\newblock Continual lifelong learning with neural networks: A review.
\newblock {\em Neural Networks}, 113:54 -- 71, 2019.

\bibitem{pflb2019comprehensive}
B.~Pfülb and A.~Gepperth.
\newblock A comprehensive, application-oriented study of catastrophic
  forgetting in {DNN}s.
\newblock In {\em ICLR 2019}, 2019.

\bibitem{PRATAMA2019150}
Mahardhika Pratama and Dianhui Wang.
\newblock Deep stacked stochastic configuration networks for lifelong learning
  of non-stationary data streams.
\newblock {\em Information Sciences}, 495:150 -- 174, 2019.

\bibitem{7737670}
R.~{Precup}, T.~{Teban}, T.~E.~A. d.~{Oliveira}, and E.~M. {Petriu}.
\newblock Evolving fuzzy models for myoelectric-based control of a prosthetic
  hand.
\newblock In {\em 2016 IEEE International Conference on Fuzzy Systems
  (FUZZ-IEEE)}, pages 72--77, 2016.

\bibitem{Rebuffi2017}
Sylvestre~Alvise Rebuffi, Alexander Kolesnikov, Georg Sperl, and Christoph~H.
  Lampert.
\newblock {iCaRL: Incremental classifier and representation learning}.
\newblock In {\em CVPR 2017}, 2017.

\bibitem{6246152}
A.~{Reiss} and D.~{Stricker}.
\newblock Introducing a new benchmarked dataset for activity monitoring.
\newblock In {\em ISWC 2012}, pages 108--109, June 2012.

\bibitem{REYESORTIZ2016754}
Jorge-L. Reyes-Ortiz, Luca Oneto, Albert SamÃ , Xavier Parra, and Davide
  Anguita.
\newblock Transition-aware human activity recognition using smartphones.
\newblock {\em Neurocomputing}, 171:754 -- 767, 2016.

\bibitem{Rios_Itti18nipscl}
A.~Rios and L.~Itti.
\newblock Closed-loop gan for continual learning.
\newblock In {\em 2018 NIPS workshop on Continual Learning}, pages 1--8, Dec
  2018.

\bibitem{ROS201386}
M.~Ros, M.P. Cuéllar, M.~Delgado, and A.~Vila.
\newblock Online recognition of human activities and adaptation to habit
  changes by means of learning automata and fuzzy temporal windows.
\newblock {\em Information Sciences}, 220:86 -- 101, 2013.
\newblock Online Fuzzy Machine Learning and Data Mining.

\bibitem{prognet}
Andrei~A. Rusu, Neil~C. Rabinowitz, Guillaume Desjardins, Hubert Soyer, James
  Kirkpatrick, Koray Kavukcuoglu, Razvan Pascanu, and Raia Hadsell.
\newblock Progressive neural networks.
\newblock {\em CoRR}, abs/1606.04671, 2016.

\bibitem{DGR}
Hanul Shin, Jung~Kwon Lee, Jaehong Kim, and Jiwon Kim.
\newblock Continual learning with deep generative replay.
\newblock In {\em Advances in Neural Information Processing Systems}, pages
  2990--2999, 2017.

\bibitem{ven2019scenarios}
van~de Ven, Gido M, and Andreas~S Tolias.
\newblock Three scenarios for continual learning.
\newblock {\em arXiv preprint arXiv:1904.07734}, 2019.

\bibitem{vanKasteren2011}
T.~L.~M. van Kasteren, G.~Englebienne, and B.~J.~A. Kr{\"o}se.
\newblock {\em Human Activity Recognition from Wireless Sensor Network Data:
  Benchmark and Software}, pages 165--186.
\newblock Atlantis Press, Paris, 2011.

\bibitem{ViroliZ10}
Mirko Viroli and Franco Zambonelli.
\newblock A biochemical approach to adaptive service ecosystems.
\newblock {\em Information Sciences}, 180(10):1876--1892, 2010.

\bibitem{WANG20193}
Jindong Wang, Yiqiang Chen, Shuji Hao, Xiaohui Peng, and Lisha Hu.
\newblock Deep learning for sensor-based activity recognition: A survey.
\newblock {\em Pattern Recognition Letters}, 119:3 -- 11, 2019.

\bibitem{Welling2009}
Max Welling.
\newblock Herding dynamical weights to learn.
\newblock In {\em Proceedings of the 26th Annual International Conference on
  Machine Learning}, page 1121–1128, New York, NY, USA, 2009. Association for
  Computing Machinery.

\bibitem{wen2018fewshot}
Junfeng Wen, Yanshuai Cao, and Ruitong Huang.
\newblock Few-shot self reminder to overcome catastrophic forgetting.
\newblock {\em ArXiv}, abs/1812.00543, 2018.

\bibitem{8954008}
Y.~{Wu}, Y.~{Chen}, L.~{Wang}, Y.~{Ye}, Z.~{Liu}, Y.~{Guo}, and Y.~{Fu}.
\newblock Large scale incremental learning.
\newblock In {\em 2019 IEEE/CVF Conference on Computer Vision and Pattern
  Recognition (CVPR)}, pages 374--382, 2019.

\bibitem{8706961}
H.~Y. {Yatbaz}, S.~{Eraslan}, Y.~{Yesilada}, and E.~{Ever}.
\newblock Activity recognition using binary sensors for elderly people living
  alone: Scanpath trend analysis approach.
\newblock {\em IEEE Sensors Journal}, 19(17):7575--7582, 2019.

\bibitem{8903481}
J.~{Ye}, S.~{Dobson}, and F.~{Zambonelli}.
\newblock Lifelong learning in sensor-based human activity recognition.
\newblock {\em IEEE Pervasive Computing}, 18(3):49--58, 2019.

\bibitem{ye2020evolving}
Juan Ye and Elise Callus.
\newblock Evolving models for incrementally learning emerging activities.
\newblock {\em Journal of Ambient Intelligence and Smart Environments},
  12:313--325, 2020.

\bibitem{ye2016discovery}
Juan Ye, Lei Fang, and Simon Dobson.
\newblock Discovery and recognition of unknown activities.
\newblock In {\em Proceedings of Ubicomp '16 Adjunct}, pages 783--792. ACM,
  2016.

\bibitem{YE201632}
Juan Ye, Graeme Stevenson, and Simon Dobson.
\newblock Detecting abnormal events on binary sensors in smart home
  environments.
\newblock {\em Pervasive and Mobile Computing}, 33:32 -- 49, 2016.

\bibitem{yoon2018lifelong}
Jaehong Yoon, Eunho Yang, Jeongtae Lee, and Sung~Ju Hwang.
\newblock Lifelong learning with dynamically expandable networks.
\newblock ICLR, 2018.

\bibitem{8796471}
K.~{Yordanova}.
\newblock Challenges providing ground truth for pervasive healthcare systems.
\newblock {\em IEEE Pervasive Computing}, 18(2):100--104, 2019.

\bibitem{conf/iccv/ZhaiCTHNM19}
Mengyao Zhai, Lei Chen, Frederick Tung, Jiawei He, Megha Nawhal, and Greg Mori.
\newblock Lifelong gan: Continual learning for conditional image generation.
\newblock In {\em ICCV}, pages 2759--2768. IEEE, 2019.

\bibitem{8645807}
Y.~{Zhang}, G.~{Tian}, S.~{Zhang}, and C.~{Li}.
\newblock A knowledge-based approach for multiagent collaboration in smart
  home: From activity recognition to guidance service.
\newblock {\em IEEE Transactions on Instrumentation and Measurement},
  69(2):317--329, 2020.

\bibitem{zhao2019maintaining}
Bowen Zhao, Xi~Xiao, Guojun Gan, Bin Zhang, and Shutao Xia.
\newblock Maintaining discrimination and fairness in class incremental
  learning.
\newblock {\em arXiv preprint arXiv:1911.07053}, 2019.

\end{thebibliography}
\bibliographystyle{plain}

\end{document}